\documentclass[aps,prl,superscriptaddress,twocolumn,a4,footinbib]{revtex4-1} 
\usepackage[pdftex]{graphicx}
\usepackage{dcolumn}
\usepackage{bm}
\usepackage{fancyhdr}
\usepackage{float}
\usepackage{ifthen}
\usepackage{eurosym}
\usepackage[usenames,dvipsnames]{color}
\usepackage{hyperref}
\usepackage{amsmath}
\usepackage{amssymb}
\newcommand{\vv}[1]{\boldsymbol{#1}}

\newcommand{\ximp}{\ensuremath{n_\mathrm{imp}}}
\newcommand{\Tc}{\ensuremath{T_\mathrm{c}}}
\newcommand{\Ts}{\ensuremath{T_\mathrm{spi}}}
\newcommand{\TN}{\ensuremath{T_\mathrm{N}}}
\newcommand{\mAF}{\ensuremath{m_\mathrm{AF}}}
\newcommand{\mindist}{\ensuremath{\xi_\mathrm{min}}}

\begin{document}

\title{Multiferroic magnetic spirals induced by random magnetic exchanges}
\author{Andrea Scaramucci}
\email{andrea.scaramucci@psi.ch}
\affiliation{Laboratory for Scientific Development and Novel Materials, Paul Scherrer Institut, 5235, Villigen PSI, Switzerland}
\author{Hiroshi Shinaoka}
\email{shinaoka@itp.phys.ethz.ch}
\affiliation{Institute for Theoretical Physics, ETH Zurich, CH-8093 Z\"{u}rich, Switzerland}
\affiliation{Department of Physics, University of Fribourg, 1700 Fribourg, Switzerland}
\author{Maxim V. Mostovoy}
\affiliation{Zernike Institute for Advanced Materials,
University of Groningen, Nijenborgh 4,
9747 AG, Groningen, The Netherlands}
\author{Markus M\"{u}ller}
\affiliation{Condensed  Matter  Theory  Group,  Paul  Scherrer  Institute,  CH-5232  Villigen  PSI,  Switzerland}
\affiliation{The Abdus Salam International Centre for Theoretical Physics, 34151, Trieste, Italy}
\author{Christopher Mudry}
\affiliation{Condensed  Matter  Theory  Group,  Paul  Scherrer  Institute,  CH-5232  Villigen  PSI,  Switzerland}
\author{Matthias Troyer}
\affiliation{Institute for Theoretical Physics, ETH Zurich, CH-8093 Z\"{u}rich, Switzerland}
\author{Nicola A. Spaldin}
\affiliation{Materials Theory, ETH Zurich, CH-8093 Z\"{u}rich, Switzerland}

\begin{abstract}
Multiferroism can originate from the breaking of inversion symmetry
caused by magnetic-spiral order.  The usual mechanism for stabilizing
a magnetic spiral is competition between magnetic exchange
interactions differing by their range and sign, such as
nearest-neighbor and next-nearest-neighbor interactions. Since the
latter are usually weak the onset temperatures for multiferroism via
this mechanism are typically low.  By considering a realistic model
for YBaCuFeO$_5$ we propose an alternative mechanism for
magnetic-spiral order, and hence for multiferroism, that occurs at
much higher temperatures.  We show using Monte-Carlo simulations and
electronic structure calculations based on density functional theory
that the Heisenberg model on a geometrically non-frustrated lattice
with only nearest-neighbor interactions can have a spiral phase up
to high temperature when frustrating bonds 
are introduced randomly along a single crystallographic direction 
as caused, e.g., by a particular type of chemical disorder. 
This long-range correlated pattern of frustration avoids 
ferroelectrically inactive spin glass order.  
Finally, we provide an intuitive
explanation for this mechanism and discuss 
its generalization to other materials.
\end{abstract}

\maketitle 

Insulators with magnetic spiral order are of particular interest
because of their associated multiferroism%
~\cite{Katsura_2005,Mostovoy_2006,khomskii_trend:_2009,Tokura_2010,Kimura_2007}
in which the breaking of inversion symmetry by the magnetic spiral
drives long-range ferroelectric order.  
The magnetic order can then be manipulated by an applied voltage with
minimal current dissipation due to the insulating nature offering
potential for low loss memory devices.

For many insulators, such as the orthorhombic manganites 
$R$MnO$_3$ ($R$=Dy,Tb)%
~\cite{kimura_magnetic_2003,goto_ferroelectricity_2004,kenzelmann_magnetic_2005,Dagotto_2008,mochizuki_microscopic_2009,mochizuki_microscopic_2009}, 
spiral order results from a competition between
nearest-neighbor and further-neighbor magnetic exchanges 
of comparable strength. 
As a consequence the onset temperature $T^{\,}_{\mathrm{spi}}$ is set
by the rather low energy scale of a typical further-neighbor
exchange, strongly limiting the multiferroic temperature range. 
Alternative routes to stabilizing magnetic
spirals at higher temperatures are, therefore, of fundamental and 
technological interest.

The phenomenology of the spiral magnet YBaCuFeO$_5$, 
which has one of the highest critical temperatures among
the magnetically driven multiferroics
~\cite{kundys_multiferroicity_2009,morin_incommensurate_2015},
suggests that a particular type of chemical disorder might 
provide such a route.
As the temperature is lowered below $T^{\,}_{\mathrm{N}} \sim 440$ K in
YBaCuFeO$_5$, the paramagnetic state undergoes a transition to a
commensurate magnetic order with wave vector
$\boldsymbol{q}^{\,}_{\mathrm{N}}=
(\frac{1}{2},\frac{1}{2},\frac{1}{2})$.
Then, below
$T^{\,}_{\mathrm{spi}}<T^{\,}_{\mathrm{N}}$, a multiferroic magnetic
spiral phase sets in, 
with a propagation wave vector along the $c$ crystallographic axis,
$\vv{q}^{\,}_{\mathrm{spi}}=(\frac{1}{2},\frac{1}{2},\frac{1}{2}-Q)$.
The value of $Q$ increases smoothly from $Q(T^{\,}_{\mathrm{spi}})=0$ as
temperature is decreased.  Importantly, the reported values of
$T^{\,}_{\mathrm{spi}}$ range from 180 K to 310~K%
~\cite{kundys_multiferroicity_2009,morin_incommensurate_2015,caignaert_crystal_1995,kawamura_high-temperature_2010,ruiz-aragon_low-temperature_1998,Morin_2016}
depending on the preparation conditions,
and it was recently shown%
~\cite{Morin_2016} that $T^{\,}_{\mathrm{spi}}$ and $Q$ increase
systematically with Fe$^{3+}$/Cu$^{2+}$ occupational disorder.
These observations suggest that chemical disorder 
plays an essential role in stabilizing the magnetic spiral motivating
our search for a microscopic mechanism by which disorder facilitates, 
or even drives, magnetic spiral order.
 
In this paper, we introduce a classical Heisenberg spin model for
YBaCuFeO$_5$ in which spiral order is indeed induced by chemical
disorder. We describe the local moments of Cu$^{2+}$ and Fe$^{3+}$
as classical Heisenberg spins, $\vv{S}^{\,}_{\vv{r}}$, 
localized at the positions $\vv{r}$ of a lattice.
We assume only nearest-neighbor exchange interactions, and use the magnitudes 
calculated from the local spin density approximation including an effective
Hubbard $U$ correction (LSDA+U) 
for YBaCuFeO$_5$%
~\footnote{
Note that the parameters used in this paper differ sligthly from those of
Ref.\, \onlinecite{morin_incommensurate_2015}
as the LSDA+U method used to extract them in
Ref.\, \onlinecite{morin_incommensurate_2015}
differs from the one used here.
          }.
Without chemical disorder, the magnet is unfrustrated and establishes
commensurate antiferromagnetic order at $T^{\,}_{\mathrm{N}}\approx$ 300
K. Frustration is introduced through dilute impurity bonds with enhanced
exchange couplings of opposite sign. The structure of YBaCuFeO$_5$ is
such that the chemical disorder introduces only collinear impurity
bonds parallel to the $c$ axis. We show that the induced
frustration results in a local canting of the antiferromagnetic order
parameter around the impurity bond, which spontaneously breaks the
local inversion symmetry around each impurity so that
$\vv{S}^{\,}_{\vv{r}^{\,}_{\mathrm{imp}}+\delta\vv{r}}\neq
\vv{S}^{\,}_{\vv{r}^{\,}_{\mathrm{imp}}-{\delta\vv{r}}}$.  
The Goldstone modes of the antiferromagnet mediate a long-range coupling
between the cantings of different impurities and establish long-range
order among the cantings that induces a continuous twist of the
antiferromagnetic order parameter in the direction parallel to the
impurity bonds. 
Related mechanisms are likely responsible for the numerical results obtained in the two-dimensional systems studied by Ivanov et {\it al.} \cite{Ivanov_square-lattice_1996} and by Capati et {\it al.} \cite{Lorenzana2015}. There, too, impurity bonds were orientationally correlated.
We note that randomly oriented impurity bonds would
have a spin glass solution
~\cite{Edwards_1975,Binder1986,villain_two-level_1977,villain_two-level_1978},
whose magnetic order does not couple to a net electric polarization
and thus does not lead to multiferroism. 
This mechanism results in a
$T^{\,}_{\mathrm{spi}}$ of the order of a typical exchange coupling.
Our Monte Carlo simulations for $\mathrm{YBaCuFeO}_5$ yield
$T^{\,}_{\mathrm{spi}}$
as high as 250~K depending on the concentration of the impurity
bonds and their strength, 
in a  manner that is consistent with the experimentally
observed dependence of $T^{\,}_{\mathrm{spi}}$ and
$\vv{q}^{\,}_{\mathrm{spi}}$ on the amount of Fe$^{3+}$/Cu$^{2+}$
occupational disorder \cite{Morin_2016}.

\begin{figure*}[ht!]
\centerline{\includegraphics[width=2\columnwidth]{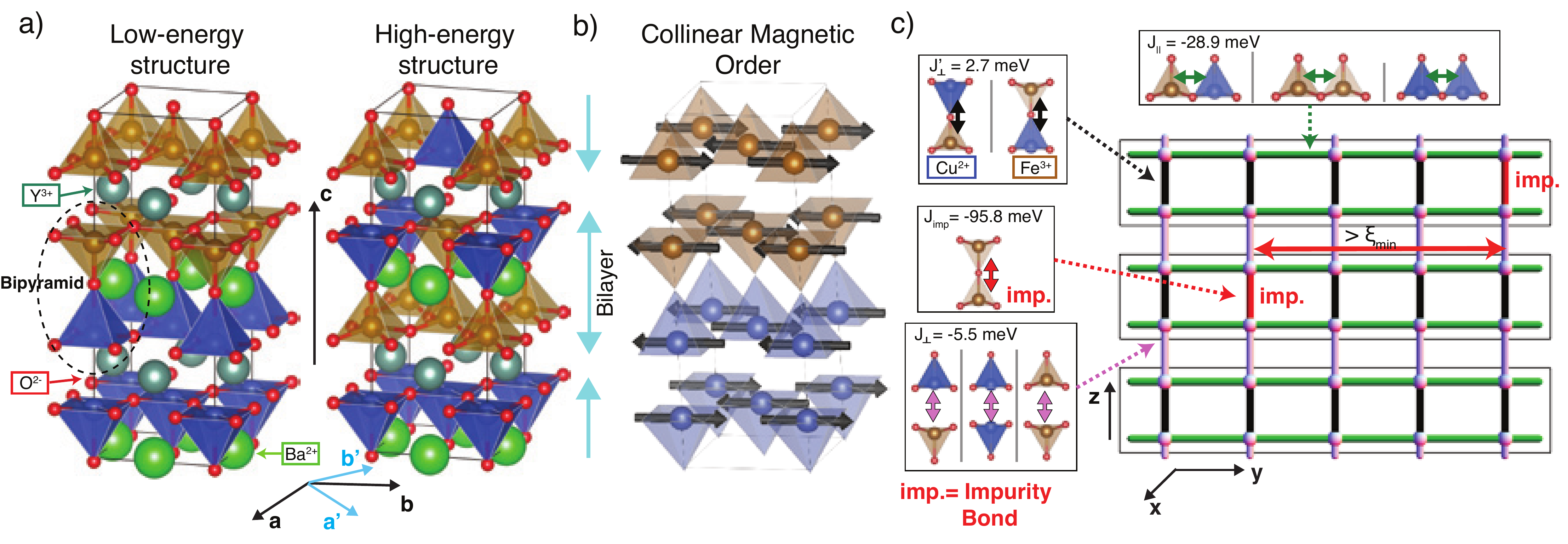}}
\caption{
(a)
Two different orderings of Cu$^{2+}$ and Fe$^{3+}$ in the 
$\sqrt{2}\times \sqrt{2}\times2$ supercell of YBaCuFeO$_5$.
The FeO$_5$ and the CuO$_5$ square pyramids are shown in gold and blue, 
respectively. The left panel shows a low-energy structure consisting only of 
Fe$^{3+}$--Cu$^{2+}$ bipyramids.
The right panel shows a high-energy structure with one Fe$^{3+}$-Fe$^{3+}$ 
and, upon periodic repetition, one Cu$^{2+}$-Cu$^{2+}$ bipyramid.
(b) Unfrustrated magnetic order in the commensurate phase.    
(c) Simplified spin model for YBaCuFeO$_5$ with nearest-neighbor exchange
interactions. The $a^{\prime}$- and $b^{\prime}$-axes are
rotated by 45$^\circ$ with respect to the $a$- and $b$-axes in (a). 
The magnetic ions are depicted as pink spheres. 
They correspond to either Cu$^{2+}$ or Fe$^{3+}$. 
They form stacked bilayers (regions enclosed by the rectangles)
obtained by stacking a pair of $x$--$y$ square lattices 
(corresponding to YBaCuFeO$_5$ bilayers) along the $z$ axis. 
The antiferromagnetic exchange coupling $J^{\,}_{\parallel}$ 
within a layer is assumed to be independent of whether it couples 
Fe$^{3+}$ with Fe$^{3+}$, Cu$^{2+}$ with Cu$^{2+}$ or Fe$^{3+}$ with Cu$^{2+}$.
The inter-layer coupling $J^{\prime}_{\perp}$ within a
bilayer is ferromagnetic for Cu$^{2+}$-Fe$^{3+}$ bipyramids (see text). 
Impurity bonds have a strong antiferromagnetic exchange, $J^{\,}_\mathrm{imp}$, 
similar in magnitude to the \textit{ab initio} value for Fe$^{3+}$-Fe$^{3+}$ 
bipyramids. The impurity bonds are randomly distributed, 
up to the constraint that two impurity bonds in
adjacent bilayers cannot be closer than a minimal distance
$\xi^{\,}_{\mathrm{min}}$.  
       }
\label{fig:Model}
\end{figure*}

\textit{Microscopic origin of spin-spiral state in $\mathrm{YBaCuFeO}_5$.}
YBaCuFeO$_5$ forms a vacancy ordered perovskite structure
for which planes of Y ions separate bilayers of BaCuFeO$_5$. 
The latter consist of
corner-sharing FeO$_5$/CuO$_5$ bipyramids,
as depicted in Fig.\ \ref{fig:Model}(a).		

A recent \textit{ab initio} study
\cite{morin_incommensurate_2015} revealed that the lowest energy
arrangements of Fe and Cu ions contain only Fe$^{3+}$--Cu$^{2+}$
bipyramids as shown in the left panel of Fig.\ \ref{fig:Model}(a). 
The exchange interactions between the magnetic ions 
making up these low-energy structures were also calculated 
using LSDA+U. The exchange interactions within 
the Fe-Cu bipyramids were found 
to be uniform in sign (ferromagnetic), 
and so unfrustrated, and thus cannot explain 
the emergence of a magnetic spiral.
Bipyramids of Cu$^{2+}$--Cu$^{2+}$ and Fe$^{3+}$--Fe$^{3+}$, 
shown in the right panel of
Fig.\ \ref{fig:Model}(a), are energetically more costly, 
but they nevertheless occur as local
defects that form during preparation and thermal treatment of the sample.
The Fe$^{3+}$--Fe$^{3+}$ bipyramids introduce strongly 
frustrating antiferromagnetic couplings along the $c$ axis%
~\footnote{
The antiferromagnetic Fe$^{3+}$--Fe$^{3+}$ coupling ranges between $90$ meV 
and $94$ meV in magnitude depending on the neighboring bipyramids.
(We use the convention that the magnetic moments $|\vv{S}|$ are 
of unit length.) 
\textit{Ab initio} calculations were done using the same method
as in Ref.\ \cite{morin_incommensurate_2015}
          }. 
In the following, we show that a small
concentration of Fe$^{3+}$-Fe$^{3+}$ bipyramids
is at the microscopic origin of the spiral order.
          
We model the magnetic ordering of YBaCuFeO$_5$ 
using the classical Hamiltonian
\begin{equation}
H=
-\frac{1}{2}
\sum_{\vv{r},\vv{r}'}   
J^{\,}_{\vv{r},\vv{r}'}\, 
\vv{S}^{\,}_{\vv{r}}\cdot \vv{S}^{\,}_{\vv{r}'} 
+ 
\frac{\Delta}{2} 
\sum_{\vv{r}} (\vv{S}^{\,}_{\vv{r}}\cdot\hat{\vv{n}})^{2}.
\label{Eq:SpinHam}
\end{equation}
Here, the classical spin $\vv{S}_{\vv{r}}$ is a three-component unit
vector located at the site $\vv{r}$ of a cubic lattice 
isomorphic to the tetragonal lattice of YBaCuFeO$_5$. 
To distinguish between the lattice vectors of the real material and
those of the spin model we label the latter as $\vv{x}$, $\vv{y}$, and
$\vv{z}$. Their length is equal to $a_{\,}^{\prime}=b_{\,}^{\prime}
\approx c/2$ where $\vv{a}^{\prime}_{\,}$, $\vv{b}^{\prime}_{\,}$, and
$\vv{c}$ are the lattice vectors of YBaCuFeO$_5$ as shown in
Fig.\ {\ref{fig:Model} (with $\vv{c}\parallel\vv{z}$).  To facilitate
comparison with experiments, we express the wave vectors in units of
the reciprocal vectors of the crystallographic unit cell of
YBaFeCuO$_5$ throughout. 
We retain only nearest-neighbor exchange interactions.  The $ab$
planes host spins on a square lattice, coupled antiferromagnetically
with the strong exchange $J^{\,}_{\parallel}=-28.9\,\mathrm{meV}$
~\footnote{ 
The magnetic exchange couplings
have been rescaled so as to account for the magnitude of the classical
magnetic moments of the ions located on the sites $\vv{r}$ and
$\vv{r}'$.  }. 
These planes are stacked along the $c$ axis in
bilayers (Fig.\ \ref{fig:Model}).  Within a
bilayer, the nearest neighbors along the $c$ axis, that is the spins in
a bipyramid, are coupled with the weak ferromagnetic exchange
interaction $J^{\prime}_{\perp}=2.7\,\mathrm{meV}>0$.  Adjacent bilayers
are coupled with the relatively weak antiferromagnetic coupling
$J^{\,}_{\perp}=-5.5\,\mathrm{meV}$. Finally, the defect bipyramids 
are modelled by a small concentration,
$n^{\,}_{\mathrm{imp}}$, of randomly located ``impurity bonds''
lying within the bilayers. They substitute
$J^{\prime}_{\perp}>0$ with the strong antiferromagnetic coupling
$J^{\,}_\text{imp}=-95.8\,\mathrm{meV}$  calculated for
Fe$^{3+}$--Fe$^{3+}$ bipyramids and locally frustrate the
ferromagnetic intra-bilayer couplings inducing a local canting for 
$|J^{\,}_\text{imp}|$
sufficiently larger than $J^{\,}_\parallel$. 
At the same time, we assume impurities to be
sufficiently dilute, 
($n^{\,}_{\mathrm{imp}}\, J^{\,}_{\mathrm{imp}}\lesssim J^{\prime}_{\perp}$), 
so that intralayer antiferromagnetic alignment does not become favored
within bilayers.
The impurity bonds are long-range correlated in the sense that they are always
oriented parallel to the $c$ axis.
We do not include the Cu$^{2+}$--Cu$^{2+}$ bipyramids 
(which stoichiometry implies to be as abundant as the Fe$^{3+}$-Fe$^{3+}$ 
bipyramids) since their inter-layer exchange is substantially smaller 
than all other couplings.

The second term in Eq.\ (\ref{Eq:SpinHam}) 
describes a small easy plane anisotropy with $\Delta =0.5$ 
meV which favors magnetic moments in the plane perpendicular to the 
direction defined by the unit vector
$\hat{\vv{n}}$. The spiral order parameter is defined by
\begin{equation}
P= 
\frac{1}{N} 
\sum_{\vv{r}} 
(\mathbf{S}^{\,}_{\vv{r}}\wedge \mathbf{S}^{\,}_{\vv{r}+\vv{z}}) 
\cdot
\hat{\vv{n}},
\label{eq: def spiral order parameter}
\end{equation} 
where $N$ is the number of spins in the lattice. 
Neutron scattering shows that $\hat{\vv{n}}\neq\hat{\vv{c}}$, 
which ensures a cycloidal component of the spiral and thus a coupling 
to the electric polarization%
~\footnote{ 
Experiment shows that $\vv{n} \neq \hat{\vv{c}}$, which 
guarantees that the spiral has a cycloidal component. 
Note that a cycloidal component is a sufficient condition
for the spiral order parameter to couple to the electric polarization.
However, it is not a necessary condition when
the impurity distribution breaks the two-fold
rotation symmetry about an axis perpendicular to the spiral wave vector. 
          }.  
For the calculation of the
thermodynamics, the orientation of $\hat{\vv{n}} $ plays no role.

In our Monte Carlo
simulations, we use a cubic superlattice of linear dimension 
$L\times L\times 2L$  (where $L\le 28$) 
with periodic boundary conditions in plane 
and open boundary conditions in the $\vv{z}$ direction.  
Impurity bonds are randomly located, but subject to the
constraint that any two impurity bonds in adjacent bilayers
are further apart than a cutoff in-plane distance $\mindist$ 
(tuned to be $2.5$ or $4$ in units of the lattice spacing).  
This cutoff embodies the main effect of the strong Coulomb
repulsion between Fe$^{3+}$--Fe$^{3+}$ bipyramids.  
We consider small impurity bond concentrations
$\ximp~\le 0.04\sim J^{\prime}_{\perp}/|J^{\,}_{\mathrm{imp}}|$ 
per unit cell.  
Monte Carlo results include a disorder averaging
over 16 configurations of the impurity bonds.  
A more complete description of the model, the method and
detailed Monte Carlo data including system-size dependence are given
in the supplemental material \cite{Supplemental}.

For the clean, unfrustrated case ($\ximp=0$),
we find a transition at $T^{\,}_\mathrm{N} \simeq 300$ K $=O(J^{\,}_\parallel)$
from the high-temperature paramagnetic phase to a low-temperature
collinear antiferromagnetic phase
with the ordering wave vector $(\frac{1}{2},\frac{1}{2},\frac{1}{2})$.
This is consistent with the antiferromagnetic ordering observed 
experimentally at high temperature in Ref.\ \cite{morin_incommensurate_2015} 
and shown in Fig.\ \ref{fig:Model}(b).  The calculated specific heat 
$C$ at constant volume is shown in the first panel of Fig.~\ref{fig:MC}(a).
It shows a typical $\lambda$-peak at $T^{\,}_\mathrm{N}$, 
characteristic of a continuous transition.

With a finite concentration of  impurity bonds, $\ximp>0$,
the peak in $C$ broadens, while its position remains almost constant 
as long as $\ximp\le 0.04$.  
However, the magnitude of the collinear order parameter, 
$\mAF$, shown in the second panel of
Fig.\ \ref{fig:MC}(a), 
 is strongly suppressed below $T=T^{\,}_{\mathrm{spi}}\simeq150$ K
for $n^{\,}_{\mathrm{imp}}=0.02$.
This fact suggests the onset of spiral order.
Simultaneously, the electric polarization 
(estimated as
$\sqrt{\langle P^{2}\rangle}$)
becomes non-zero. The associated
susceptibility $\chi^{\,}_{P}$ exhibits a peak,
which seems to diverge as the system size increases,
as shown in the supplemental material \cite{Supplemental}.
In Fig.\ \ref{fig:MC}(b), we show the spin-structure factor $S(q^{\,}_c)$ 
as a function of the wave vector $q^{\,}_{c}$ along the $c$ axis 
(averaged over $q^{\,}_{a}$ and $q^{\,}_{b}$) and temperature, 
for four values of $\ximp$.
At $\ximp=0.02$, the propagation wave vector $q^{\,}_{c}$ of
the magnetic order decreases smoothly from $\pi$ below $\Ts$,
suggesting a continuous transition from the antiferromagnetic phase 
to a spiral-ordered phase,
consistent with the experimental observations
reported in Ref.\ \cite{morin_incommensurate_2015}.
A small residual $\mAF$ below $\Ts$ remains. This might be due to either
finite-size effects or to a coexistence of the spiral and
antiferromagnetic order \cite{Supplemental}.
Figure \ref{fig:MC}(c) shows $T^{\,}_{\mathrm{spi}}$,
estimated from the peak of  $\chi^{\,}_{P}$ (see Supplemental Material),
as a function of impurity concentration.
At large impurity concentration, $n_{\rm imp} \gtrsim 0.04$, a direct transition from the paramagnet to an incommensurate spiral state with $q_{\rm spi}<1/2$ is also compatible with the finite resolution of our data.

We carried out Monte Carlo simulations for two values of the minimal
in-plane distance between impurity bonds $\mindist=2.5$ and
$\mindist=4$ for each value of $n^{\,}_{\mathrm{imp}}$.  As $\ximp$
increases, $T^{\,}_{\mathrm{spi}}$ increases and almost reaches
$T^{\,}_{\mathrm{N}}$ (as estimated from the peak in specific heat) at
$\ximp=0.04$.  Note that at $\ximp=0.04$, $S(q^{\,}_c)$ has a maximum
at $q^{\,}_c~<\pi$ even at high temperature. This behavior does not
rule out a direct transition from the paramagnetic to the spiral phase
for larger (but not too large) values of $\ximp$.  Note that our
values of $T^{\,}_{\mathrm{spi}}$ constitute lower bounds due to
finite size effects in the Monte Carlo simulations.  Indeed, the peak
values for $\chi^{\,}_P$ are still moving to higher temperatures for
the largest linear system sizes $L$ that were used in the Monte Carlo
simulations.  Hence, the finite size effects on
$T^{\,}_{\mathrm{spi}}$ are expected to be sizable for the values of
$L$ used in our Monte Carlo simulations.

Finally, we discuss the short range repulsion among the impurity bonds,
which we model by imposing $\mindist$.  We find that when we
neglect to include this effect, by setting $\mindist=0$, the low
temperature state does not support any spiral order over the range
$0.02\leq\ximp\leq0.04$ that we studied.  Instead, a ``fan state'' in
which the intralayer ferromagnetic order oscillates is stabilized, as
shown in panel (c) of Fig.~\ref{fig:3DSystems}; this state will be
discussed in more detail elswehere \cite{Scaramucci_2016_Long}.
Comparing the phase diagrams obtained for $\mindist=4$ and
$\mindist=2.5$ in Fig.\ \ref{fig:MC}, we see that $\Ts$ decreases with
decreasing $\mindist$.  
Indeed, the spiral order is favored if the presence of impurity bonds
is suppressed in directions forming small angles with the $c$ axis,
where the coupling has antiferromagnetic sign.

\begin{figure}[t!]
    \begin{tabular}{cc}

       \begin{minipage}{0.5\hsize}
       	\centerline{\includegraphics[width=\textwidth]{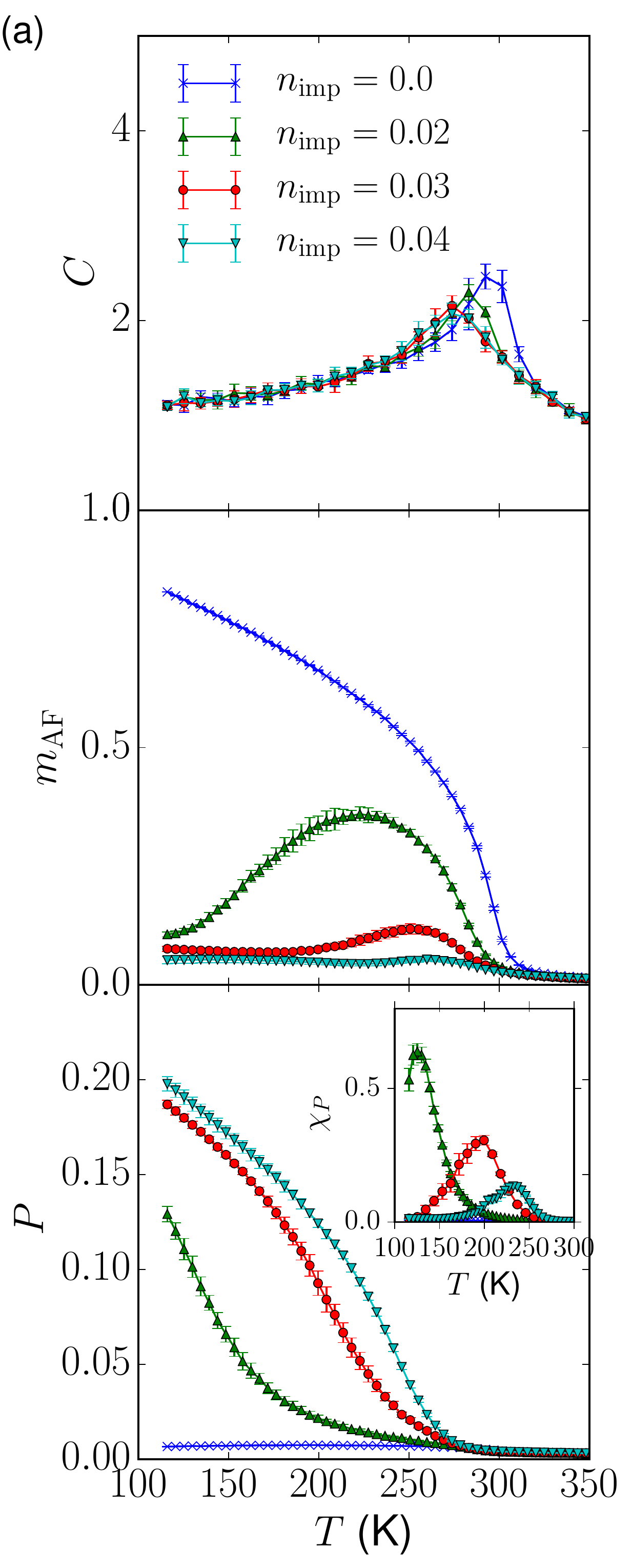}}
       \end{minipage}       
		&
       \begin{minipage}{0.5\hsize}
       \begin{tabular}{c}
       \includegraphics[width=\textwidth]{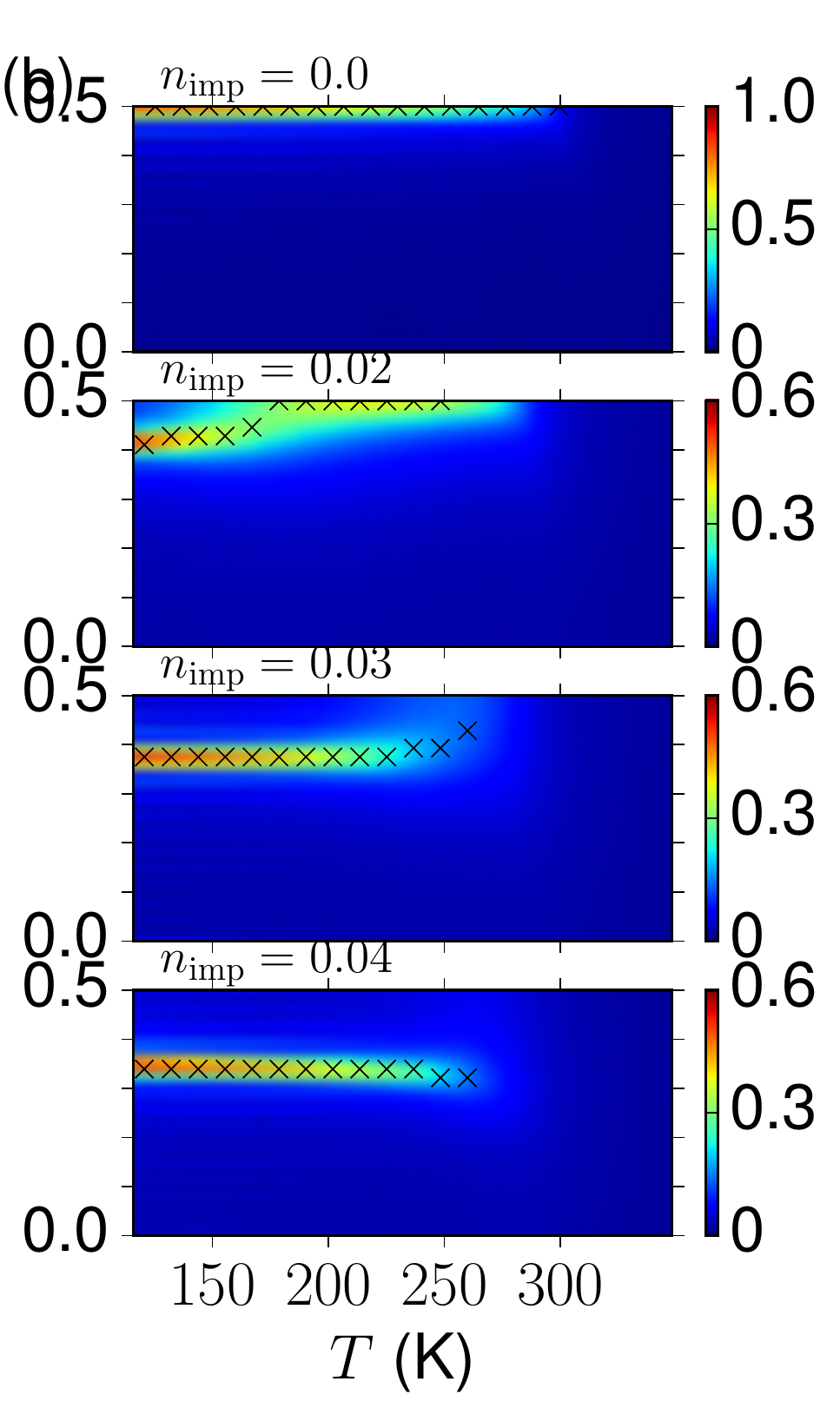}\\       
       \includegraphics[width=\textwidth]{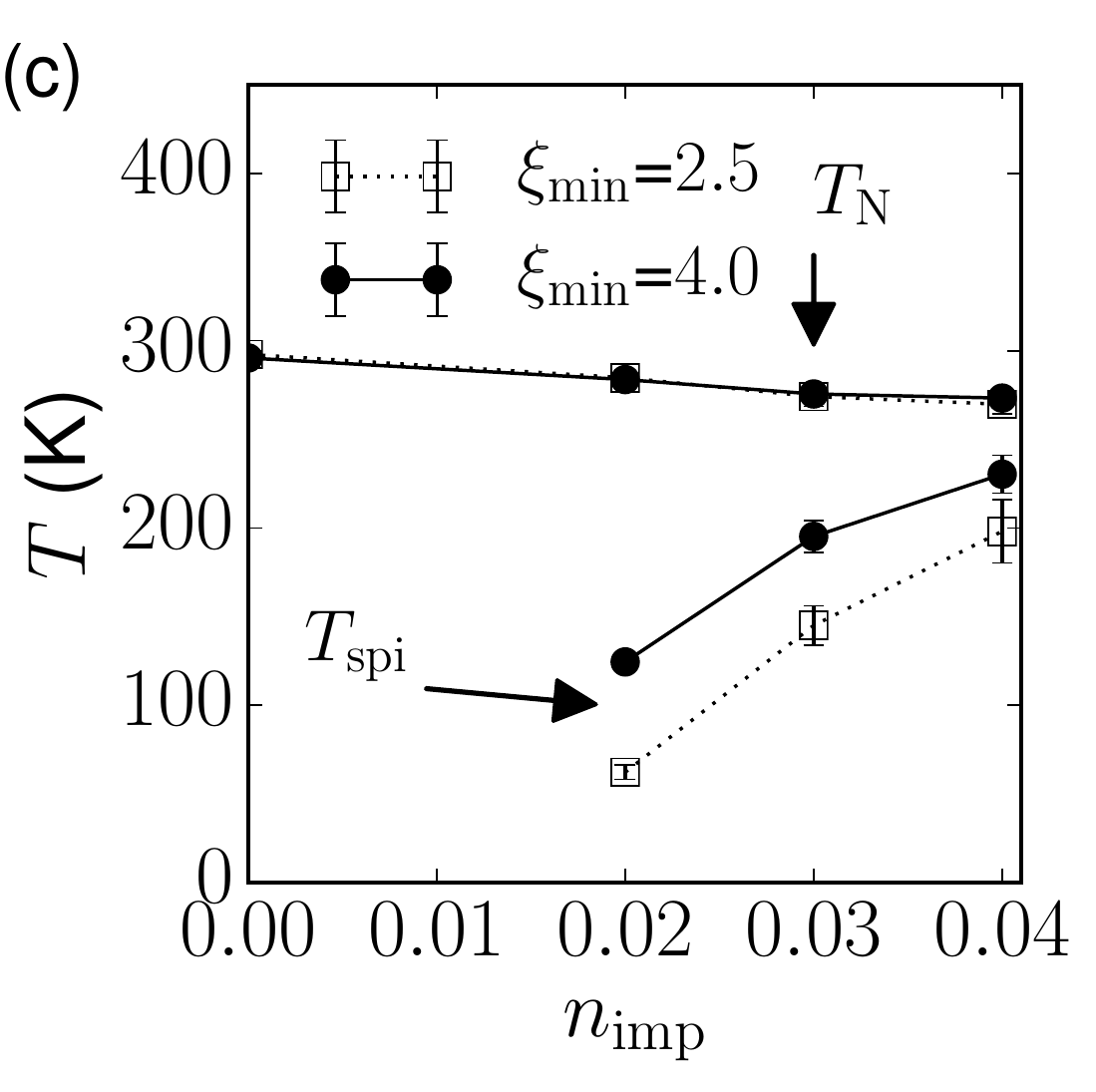}
       \end{tabular}
       \end{minipage}       
    \end{tabular}
\caption{ 
Monte Carlo results obtained for systems of size $L=28$.
(a) Specific heat $C$, collinear antiferromagnetic order parameter $\mAF$ 
and polarization $P$, which we estimated by computing
$\sqrt{\langle P^{2}\rangle}$,
for $\mindist=4$ and a range of impurity bond concentrations. 
(b) Square root of the spin-structure factor $S(q^{\,}_{c})$ computed for
$\mindist=4$. The data are averaged over the in-plane wave vectors.
The crosses represent the wave vector where $S(q_c)$ takes a maximum value at the given temperature.
The data for $n_{\rm imp}=0.04$ is compatible with a direct transition from the paramagnetic state into an incommensurate magnetic state with $q_\mathrm{spi}<1/2$.
(c) Phase diagrams obtained for $\mindist=2.5$ and $\mindist=4$. 
}
\label{fig:MC}
\end{figure}

In summary, our simulations reveal that in the range $0.02\leq\ximp\leq0.04$ 
the low temperature state is a spiral with a 
temperature-dependent wave vector 
provided the impurity bonds obey the condition that 
$\mindist\gtrsim 2.5$.
Both the spiral ordering temperature $T^{\,}_{\text{spi}}$
and the component
$1/2-\vv{q}^{\,}_{\mathrm{spi}}\cdot\hat{\vv{c}}$
of the ordering wave vector
depend on the concentration $n^{\,}_{\mathrm{imp}}$ of
the impurity bonds and are proportional 
to the latter in the limit of small $n^{\,}_{\mathrm{imp}}$.

\begin{figure*}[ht!]
\centerline{\includegraphics[width=1.9\columnwidth]{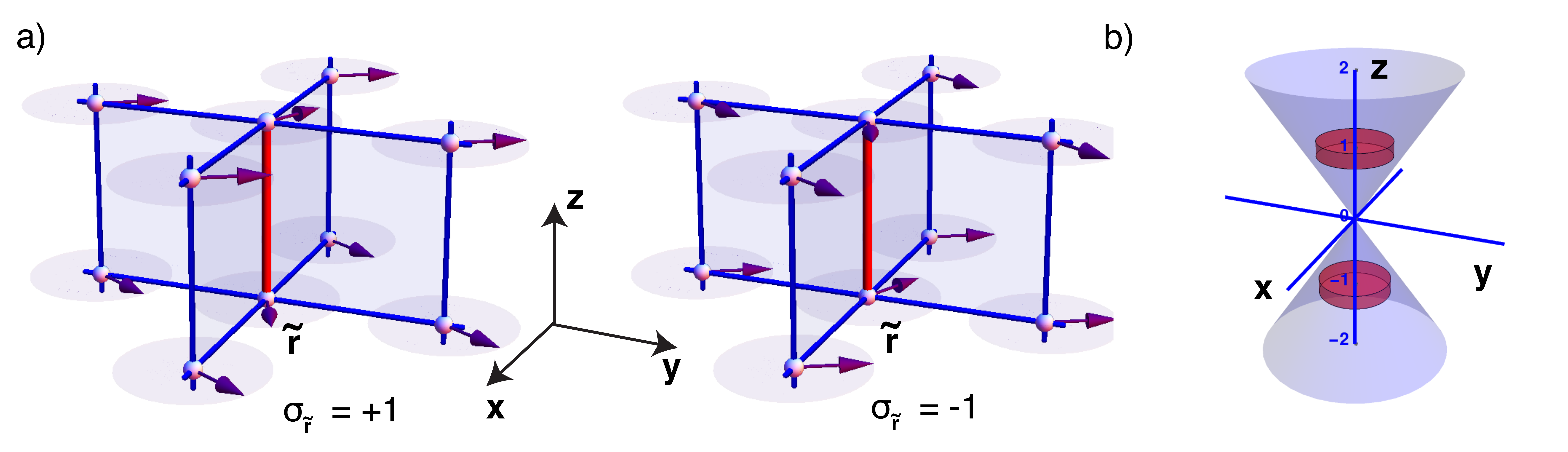}}
\caption{
(a) Sketch of the two ground states of Hamiltonian
(\ref{Eq:SpinHam}) for a single bilayer,  
in the presence of a single frustrating impurity bond (red line).
For visualization purposes the spins on 
every other site in the $ab$-plane have been reversed
so that the unfrustrated ground state looks ferromagnetic. 
The two ground states break the local inversion symmetry 
and correspond to counter-clockwise (left panel)
and clockwise (right panel) rotation of the spins 
as one proceeds along the $c$ axis. 
They can be labeled by an Ising variable
$\sigma^{\,}_{\tilde{\vv{r}}}$.  
Spin waves mediate an effective interaction 
$\Gamma^{\,}_{\tilde{\vv{r}},\tilde{\vv{r}}^{\prime}}$ 
between the local orientations
$\sigma^{\,}_{\tilde{\vv{r}}}$ and $\sigma^{\,}_{\tilde{\vv{r}}^{\prime}}$
of the cantings around two impurity bonds 
$\langle\tilde{\vv{r}},\tilde{\vv{r}}+\vv{z}\rangle$ 
and
$\langle\tilde{\vv{r}}^{\prime},\tilde{\vv{r}}^{\prime}+\vv{z}\rangle$, 
as given by Eq.\ (\ref{eq: def Ising Hamiltonian}).
(b) The blue cone is the domain of 
$\Delta\tilde{\vv{r}}=\tilde{\vv{r}}-\tilde{\vv{r}}^{\prime}$ 
for which the effective interaction
$\Gamma^{\,}_{\tilde{\vv{r}},\tilde{\vv{r}}^{\prime}}$ 
is antiferromagnetic, while outside it is ferromagnetic.
The short range constraint $\xi^{\,}_{\mathrm{min}}$ excludes the
domains represented by the red cylinders and thus increases the 
probability of a ferromagnetic coupling between neighboring Ising variables.
       }
\label{fig:Dipole}
\end{figure*}

\textit{ Mechanism for spiral stabilization -- }
Finally, we analyze the limit of low concentration  
of impurity bonds $n^{\,}_{\mathrm{imp}}$ and low temperature $T$, 
to provide insight into the mechanism
for stabilization of the spiral. 
As discussed in detail in Ref.\ \cite{Scaramucci_2016_Long}, 
impurity bonds induce cantings which are coupled through the spin
stiffnes of the hosting spin lattice.  For systems with continuous
symmetry such effective coupling is dipolar and analogous to the
Coulomb coupling that occurs for vortices in the two dimensional $XY$
model.  Moreover, the net value of dipoles couples linearly to the
winding of the spins of the hosting lattice along the $z$
axis. Therefore, distributions of impurity bonds having a ground state
with a non-vanishing net dipole stabilize a spin spiral. 
At $T\ll\Delta$, the spins become coplanar $XY$ spins with
$\vv{S}\cdot\hat{\vv{n}}=0$. For $n^{\,}_{\mathrm{imp}}=0$, the
antiferromagnetic ground state has all spins parallel in the
$xy$-plane, e.g., parallel to the $x$ axis.  
A single impurity bond with exchange of magnitude
$|J^{\,}_{\mathrm{imp}}|$ above a threshold value $J^{\,}_{\mathrm{c}}$
(where 
$J^{\,}_{\mathrm{c}} \approx J^{\,}_{\parallel} / [ C -
\ln( J^{\prime}_{\perp}/ J^{\,}_{\parallel} )/(2 \pi) ]$ with
$C\sim0.4$ for $J^{\prime}_{\perp}/J^{\,}_{\parallel}\ll1$) renders
the ground state two-fold degenerate, see Fig.\ \ref{fig:Dipole}.
Far away from the impurity bond, the staggered magnetization parallel
to the $x$ axis is restored. Close to the impurity bond, the coplanar
spins are canted away from the $x$ axis with the two ground states
differing in the local sense of rotation of spins, as given by $
P^{\,}_{\mathrm{loc}}(\tilde{\vv{r}}) = \left(
\mathbf{S}^{\,}_{\tilde{\vv{r}}}\wedge\mathbf{S}^{\,}_{\tilde{\vv{r}}+\hat{\vv{z}}}
\right) \cdot \hat{\vv{n}} $.  At finite but small
$n^{\,}_{\mathrm{imp}}>0$ the impurity bonds are well separated. Thus,
the low energy configurations can be labelled by local Ising variables
$\sigma^{\,}_{\tilde{\vv{r}}}= \mathrm{sgn}\,
P^{\,}_{\mathrm{loc}}(\tilde{\vv{r}})$ describing the sign of the
local cantings, while the remaining degrees of freedom are well
captured by spin waves.  Integrating them out yields an effective
Hamiltonian $H^{\,}_{\mathrm{eff}}$ for the Ising degrees of freedom
\begin{subequations}
\label{eq: def Ising Hamiltonian}
\begin{equation}
H^{\,}_{\mathrm{eff}}=
- 
\frac{\alpha^{2}}{2}\,
\left(
\sum_{\tilde{\vv{r}},\tilde{\vv{r}}^{\prime} \in \mathcal{L}} 
\Gamma^{\,}_{\tilde{\vv{r}},\tilde{\vv{r}}^{\prime}}\,
\sigma^{\,}_{\tilde{\vv{r}}}\,
\sigma^{\,}_{\tilde{\vv{r}}^{\prime}} 
+ 
\frac{N\,|J^{\,}_{\perp}|}{\alpha^{2}}\,Q^{2}
\right),
\label{eq: def Ising Hamiltonian a}
\end{equation}
for a system with $N$ lattice sites, where we simplified the model slightly by setting 
$|J^{\,}_{\perp}|=J^{\prime}_{\perp}$. 
We label the impurity bonds by the site $\tilde{\vv{r}}$ at their
lower end and by $\mathcal{L}$ the ensemble of such sites for a given
distribution of impurity bonds.  Here, $Q$ is the saddle-point value
of the spiral wave vector and is proportional to the net value of the
Ising spins,
\begin{equation}
Q=
-
\frac{|\alpha|}{N\,|J^{\,}_{\perp}|} 
\sum_{\tilde{\vv{r}} \in \mathcal{L}} \sigma^{\,}_{\tilde{\vv{r}}}.
\label{eq: def Ising Hamiltonian c}
\end{equation}
It is non-zero provided the ground state of the effective Ising model has a net 
magnetization.
In Eq.\ (\ref{eq: def Ising Hamiltonian a}), 
$\alpha=(|J^{\,}_{\mathrm{imp}}|+|J^{\,}_{\parallel}|)\,|P^{\mathrm{imp}}_{\mathrm{loc}}|$
where $P^{\mathrm{imp}}_{\mathrm{loc}}$ encodes
the canting angle at an isolated impurity bond \cite{Scaramucci_2016_Long}. 
As shown in Ref.\ \cite{Scaramucci_2016_Long}, at large distances
the kernel $\Gamma$ takes the (anti-) dipolar form 
\begin{equation}
\Gamma^{\,}_{\tilde{\vv{r}},\tilde{\vv{r}}^{\prime}}  
\sim  
\frac{\sqrt{|J^{\,}_{\parallel}|}}{4 \pi} 
\frac{
|J^{\,}_{\perp}|  
\left(\Delta\tilde{r}^{2}_{x}+\Delta \tilde{r}^{2}_{y}\right)
- 
2|J^{\,}_{\parallel}|\, 
\Delta\tilde{r}^{2}_{z}
     }
     {
\left[
|J^{\,}_{\perp}|
\left( 
\Delta\tilde{r}^{2}_{x} 
+ 
\Delta\tilde{r}^{2}_{y} 
\right)
+ 
|J^{\,}_{\parallel}|\, 
\Delta\tilde{r}^{2}_{z}
\right]^{\frac{5}{2}}
}.
\label{eq: def Ising Hamiltonian b}
\end{equation}
\end{subequations}

Note that $\Gamma^{\,}_{\tilde{\vv{r}},\tilde{\vv{r}}^{\prime}}  $ is ferromagnetic if
$
\Delta\tilde{r}^{2}_{x}+\Delta\tilde{r}^{2}_{y}>
\sqrt{2} |J^{\,}_{\parallel}|\,\Delta\tilde{r}^{2}_{z}/|J^{\,}_{\perp}|$ 
and antiferromagnetic otherwise, as illustrated in Fig.\ \ref{fig:Dipole}(b).
In particular, the ratio of the in-plane nearest-neighbor interaction
to the inter-plane one scales like 
$(J^{\,}_{\parallel}/J^{\,}_{\perp})^{3/2}\gg 1$.
We thus expect ferromagnetic order in-plane. The inter-layer
order might be either ferromagnetic or antiferromagnetic.
Now, the Coulomb repulsion between Fe$^{3+}$/Fe$^{3+}$ bipyramids
amounts to a suppression of antiferromagnetically coupled pairs of 
nearest-neighbor impurities. 
A mean-field calculation shows that this constraint stabilizes
ferromagnetic inter-layer order.  In the numerical simulation, we
retain the essential part of the short-distance repulsion by
forbidding impurity bonds on adjacent bilayers to be closer than
$\xi^{\,}_{\mathrm{min}}$ in plane.  Increasing
$\xi^{\,}_{\mathrm{min}}$ indeed enhances the tendency toward a
ferromagnetic Ising ground state with a finite $Q = n^{\,}_{\rm imp}
|\alpha/J_\perp|$ (as opposed to competing inter-layer
antiferromagnetic order, which has vanishing $Q$).

The latter immediately translates into a spiral order of the spins $\vv{S}^{\,}_{\vv{r}}$, with the characteristic wave vector
$\vv{q}^{\,}_{\mathrm{spi}}=(\frac{1}{2},\frac{1}{2},\frac{1}{2}-Q)$.  
The linear dependence of $Q$ on $n^{\,}_{\mathrm{imp}}$
is in qualitative agreement with the $n^{\,}_{\mathrm{imp}}$-dependence of the peak of the low $T$ structure factor, as calculated with Monte Carlo simulations and presented
in Fig.\ \ref{fig:MC} (b).

The ground state of the effective Ising  Hamiltonian 
(\ref{eq: def Ising Hamiltonian}) 
can be solved analytically when the set of impurity bonds forms a
superlattice.
Following Ref.\ \cite{Scaramucci_2016_Long}, we assume that
the impurity bonds form a Bravais lattice with the basis
$\vv{A}$, $\vv{B}$, and $\vv{C}$%
~\footnote{
We choose $\vv{A}$, $\vv{B}$, and $\vv{C}$ such that the impurity bonds are 
confined to the bilayers of the host lattice as in YBaCuFeO$_5$.
          }, 
such that $n^{\,}_{\mathrm{imp}}=
\frac{\vv{a}\cdot(\vv{b}\wedge\vv{c})}{\vv{A}\cdot(\vv{B}\wedge\vv{C})}\ll 1$.
In Figs.\ \ref{fig:3DSystems}(a) and \ref{fig:3DSystems}(b) we compare
the ground state of Eq.\ (\ref{eq: def Ising Hamiltonian a})
with the ground state obtained by Monte Carlo simulations of
Eq.\ (\ref{Eq:SpinHam}). 
For both calculations we  take $|J^{\prime}_{\perp}|=|J^{\,}_{\perp}|=4.1$ meV
to be  the average of the actual $|J^{\,}_{\perp}|$ and $|J^{\prime}_{\perp}|$ 
given in Fig.\ \ref{fig:Model}. 
In agreement with the magnetic structure obtained by the refined
analysis of the elastic neutron scattering data, the obtained magnetic
spiral has the feature that the rotation of magnetic moments mainly
happens between neighboring $ab$-layers coupled by impurity bonds
(e.g., bilayers of the YBaCuFeO$_5$ structure).
 
Fig.~\ref{fig:3DSystems}(c) is a similar comparison, but for
a superlattice favoring a state with no net winding of the spins,
i.e., a fan-like magnetic state. 

Finally, Fig.\ \ref{fig:3DSystems}(d) shows the  concentration dependence
of the  polarization $P$ 
(see Eq.\ \ref{eq: def spiral order parameter}) calculated both numerically 
from Eq.\ (\ref{Eq:SpinHam}) 
and using the effective Ising Hamiltonian (\ref{eq: def Ising Hamiltonian})
for various superlattices
for which the ground state is a spiral.
The values of polarization predicted by the effective model match well
the values of polarization obtained numerically for the ground state
of a finite system. Furthermore, the value of $P$ increases
proportionally to $n_{\mathrm{imp}}$ for small values of $P$ as
predicted by the effective model.

\begin{figure*}[ht!]
\centerline{\includegraphics[width=2\columnwidth]{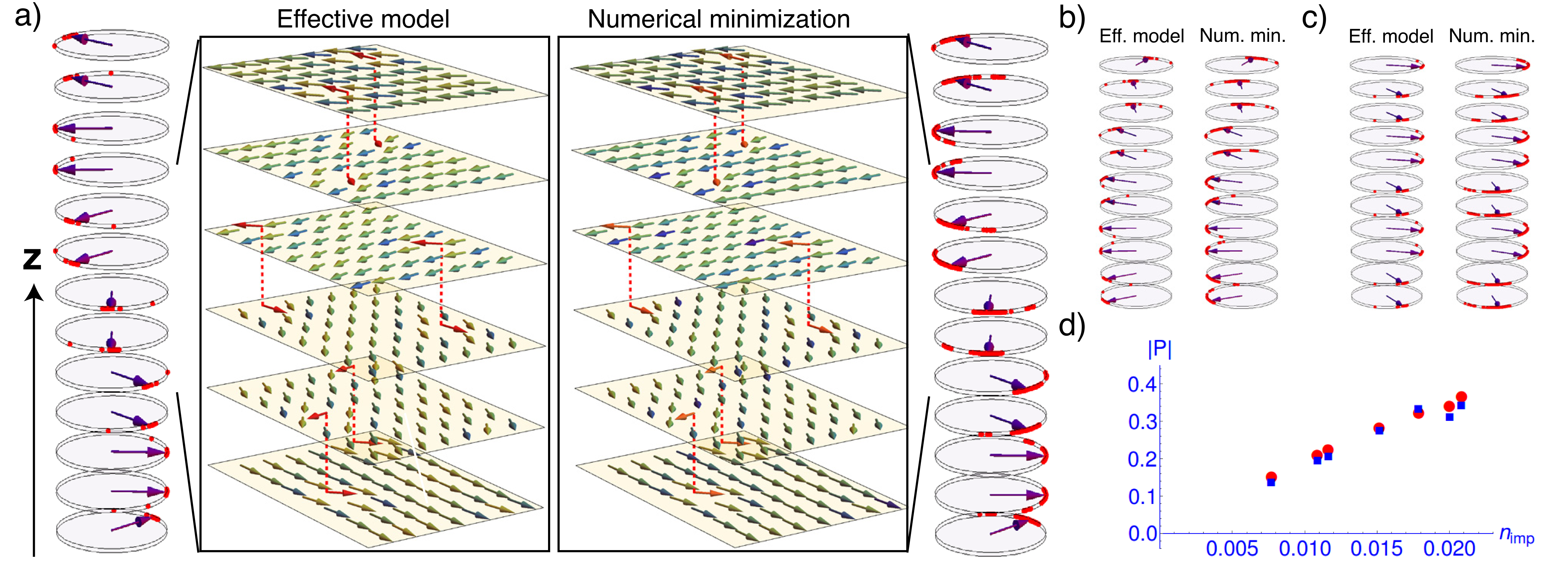}}
\caption{
(a) Comparison of a spiral ground state corresponding to an Ising ferromagnetic
ground state of 
the effective Ising Hamiltonian (\ref{eq: def Ising Hamiltonian a}) (left) 
against the spiral ground state of the 
Heisenberg Hamiltonian (\ref{Eq:SpinHam}) (right).  
The violet arrows represent the
average directions of the staggered magnetization for each layer.
The red dots lying on the circle represent the direction of each
spin in a given plane. The staggered magnetization of layers
corresponding to even bilayers have been reversed for visualization
purposes. The two central panels depict the ground states in a portion
of six layers of the lattice of overall dimensions $14\times14\times32$.
Here, the color of the arrows indicate the angular
difference of each spin from the average staggered magnetization 
of the layer and the vertical red lines depict the impurity bonds. 
For visualization purposes, spins at every other sites in the $ab$ plane 
and at every other bilayer have been reversed.  
The results were obtained for a
superlattice of impurity bonds with lattice constants
$\vv{A}=(4,3,0)$, $\vv{B}=(0,4,2)$, and $\vv{C}=(4,0,2)$.
The values of the couplings are
$J^{\,}_{\parallel}=28.9$ meV ,
$J^{\,}_{\perp}=J^{\prime}_{\perp}=4.1$ meV
and
$J^{\,}_{\mathrm{imp}}=-95.8$ meV.
(b) Comparison of the spiral ground state obtained using 
the effective Ising Hamiltonian (\ref{eq: def Ising Hamiltonian a}) (left) 
against the spiral ground state of
the Heisenberg Hamiltonian (\ref{Eq:SpinHam}) (right)
for the same value of the couplings but with 
$\vv{A}=(5,3,2)$, $\vv{B}=(3,0,4)$, and $\vv{C}=(2,4,6)$.  
Panel (c) shows the same comparison for the choice: $\vv{A}=(5,0,0)$,
$\vv{B}=(0,5,0)$, and $\vv{C}=(0,1,2)$. For such a superlattice
of impurity bonds a ``fan state'' (with no net winding along the
$c$ axis) is stabilized instead of a spiral (d) Concentration
dependence of the magnitude of $P$ defined in Eq.\ 
(\ref{eq: def spiral order parameter}) 
for the ground states obtained from the
effective Ising Hamiltonian (\ref{eq: def Ising Hamiltonian a}) (red
dots) and the Heisenberg Hamiltonian (\ref{Eq:SpinHam}) (right) (blue
squares) for various superlattices of impurity bonds stabilizing a
spiral state.  
        }
\label{fig:3DSystems}
\end{figure*}

\textit{Conclusions.}
We have presented a mechanism in which impurity bonds, when
sufficiently strong, induce a local frustration in an otherwise
non-frustrated lattice of classical spins. The associated cantings
become long-range correlated at low temperature, resulting in a spiral
magnetic order in which the sense of rotation of the spiral is
spontaneously chosen.  The spiral formation is in contrast to the spin-glass
phase expected for white-noise correlated random magnetic exchange
interactions
\cite{Edwards_1975,villain_theory_1975,villain_two-level_1977,villain_two-level_1978,Binder1986},
and is caused by orientational long-range correlations, which align
the impurity bonds along one crystallographic direction
of the lattice.  The critical temperature below which the spiral order
develops is controlled by nearest-neighbor exchange couplings, which
can be sizable, so the mechanism is relevant for engineering high
temperature multiferroics.

When the model is applied to YBaCuFeO$_5$ with realistic couplings, 
it predicts a multiferroic phase that is consistent 
with several distinct features observed experimentally.
(i) It captures the transition to a commensurate
antiferromagnet phase 
with wave vector 
$\vv{q}^{\,}_{\mathrm{N}}\equiv(\frac{1}{2},\frac{1}{2},\frac{1}{2})$
at the N\'eel temperature $T^{\,}_{\mathrm{N}}$, 
followed by a transition to an incommensurate spiral phase
with wave vector 
$\vv{q}^{\,}_{\mathrm{spi}}\equiv(\frac{1}{2},\frac{1}{2},\frac{1}{2}-Q)$
at the lower critical temperature
$T^{\,}_{\mathrm{spi}}$.
(ii) It gives rise to a magnetic spiral phase, 
in which the rotation of the magnetic moments neighboring along $c$ 
occurs mainly when they belong to the same bilayer.
(iii) The observed temperature dependence of 
$Q$ in YBaCuFeO$_5$ is well reproduced.
(iv) Finally, the dependence of both
$T^{\,}_{\mathrm{spi}}$
and 
$\vv{q}^{\,}_{\mathrm{spi}}$
on the concentration $n^{\,}_{\rm imp}$ 
is in qualitative agreement  with the dependence
on the annealing conditions of the YBaCuFeO$_5$ samples: 
The faster the quench (
and thus the higher the expected defect concentration),  
the larger the measured $T^{\,}_{\mathrm{spi}}$ and $\vv{q}^{\,}_{\mathrm{spi}}$.

We close by mentioning other compounds with magnetic spiral 
order whose origin is not understood so far, but to which the
mechanism presented in this paper might apply. 
For solid solutions of Cr$_2$O$_3$ and
Fe$_2$O$_3$ \cite{cox_magnetic_1963} 
and for the doped hexaferrite
Ba$_{(1-x)}$Sr$_x$Zn$_2$Fe$_{12}$O$_{22}$
\cite{utsumi_superexchange_2007}, the wave vector of the spiral order 
is known to change smoothly from a commensurate value at high
temperature to an incommensurate value at low temperature.  In all
cases, the maximal value of the concentration dependent
$T^{\,}_{\mathrm{spi}}$ is high ($T^{\mathrm{max}}_{\mathrm{spi}} = 148$ K
for solid solutions of Cr$_2$O$_3$ and
Fe$_2$O$_3$\cite{cox_magnetic_1963} and
$T^{\mathrm{max}}_{\mathrm{spi}}$ higher than $294$ K for
Ba$_{(1-x)}$Sr$_x$Zn$_2$Fe$_{12}$O$_{22}$ \cite{Momozawa_1985}).  
Moreover, the wave vector and the transition temperature to the
spiral state are dependent on the dopant concentration. Since in both
compounds cation substitution can introduce impurity bonds, they are
likely candidates to realize ``spiral order by disorder''.

\section{Acknowledgments}
This research was partially supported by NCCR MARVEL, funded by the
Swiss National Science Foundation.  H.S. acknowledges support from the
DFG via FOR 1346, the SNF Grant 200021E-149122, ERC Advanced Grant
SIMCOFE, ERC Consolidator Grant CORRELMAT (project number 617196).
Computer time was provided through a grant from the Swiss National
Supercomputing Centre (CSCS) under project ID p504 as well as by the
Brutus cluster of ETH Z\"{u}rich. We thank M. Kenzelmann, M. Medarde
and M. Morin for very useful discussions.

\bibliography{references,references_sup}
\pagebreak
\widetext
\begin{center}
\textbf{\large Supplemental Materials: Multiferroic magnetic spirals induced by random magnetic exchanges}
\end{center}
\setcounter{equation}{0}
\setcounter{figure}{0}
\setcounter{table}{0}
\makeatletter
\renewcommand{\theequation}{S\arabic{equation}}
\renewcommand{\thefigure}{S\arabic{figure}}

\begin{center}
The algorithm used for the Monte Carlo simulations 
is described in more details. First, we describe the model
and the way in which thermodynamic averages are computed.
Second, we compare the results obtained for the largest size of the system
considered when the minimal distance $\xi^{\,}_{\mathrm{min}}$ is set to $4$
and $2.5$ in units of in-plane lattice vectors.  Finally, 
the dependence of the thermodynamic averages on the system size 
is discussed.
\end{center}
\section{Monte Carlo (MC) simulation}
\label{sec: Monte Carlo (MC) simulation}

We model the magnetic properties of
YBaCuFeO$_5$ (YBCF) as follows.
We use the \textit{crystallographic} unit cell labelled by the vector
$\vv{r}\in\mathbb{R}^{3}$ and comprising two magnetic ions
aligned along the $\vv{c}$ direction and belonging to the same
bilayer.  Therefore, in what follows, the position of each unit cell
is defined by 
\begin{subequations}
\label{Eq:HamiltonianYBCFO}
\begin{equation}
\vv{r}:= 
m\,
\vv{a}^{\prime} 
+ 
n\,
\vv{b}^{\prime} 
+ 
2l\,
\vv{c} 
\label{Eq:HamiltonianYBCFO aa}
\end{equation}
where $m,n,l\in\mathbb{Z}$ and 
$\vv{a}^{\prime}$, 
$\vv{b}^{\prime}$, 
and $2\vv{c}$ is a basis for the
crystallographic lattice vectors.  
We distinguish the two inequivalent magnetic sublattices
with the index $\mu=1,2$.  As described in the main text, the 
classical spins $\vv{S}^{\,}_{\mu,\vv{r}}$ are unit vectors in
$\mathbb{R}^{3}$ and a single-ion easy-plane anisotropy with the
characteristic energy scale $\Delta>0$ is included.  
All together, the classical Hamiltonian 
[Eq.\,(1) in the main text] is
\begin{equation}
H^{\,}_{\mathcal{L}}:= 
- 
\frac{1}{2} 
\sum_{\substack{\vv{r},\vv{r}^{\prime}\\\mu,\nu=1,2}}
J^{\mu,\nu}_{\vv{r},\vv{r}^{\prime}}\, 
\vv{S}^{\,}_{\vv{r},\mu} 
\cdot 
\vv{S}^{\,}_{\vv{r}^{\prime},\nu} 
+ 
H^{\,}_{\mathrm{imp}}
+ 
\frac{\Delta}{2}
\sum_{\substack{\vv{r}\\ \mu = 1,2}} 
\left(S^{c}_{\vv{r},\mu}\right)^{2}.
\label{Eq:HamiltonianYBCFO a}
\end{equation}
There are three characteristic energy scales. 
The antiferromagnetic in-plane exchange coupling $J^{\,}_{\parallel}\leq0$, 
the intra-unit cell (intra-layer along the $\vv{c}$ direction) 
ferromagnetic exchange coupling $J^{\prime}_{\perp}\geq0$, 
and the inter-unit cell (inter-layer)
antiferromagnetic exchange coupling $J^{\,}_{\perp}\leq0$
(also along the $\vv{c}$ direction).
They enter Eq.\ (\ref{Eq:HamiltonianYBCFO a}) according to the rules,
\begin{align}
&
J^{1,2}_{\vv{r},\vv{r}^{\prime}}\:= 
J^{\prime}_{\perp}\,
\delta^{\,}_{\vv{r},\vv{r}^{\prime}}+ J^{\,}_{\perp} \delta^{\,}_{\vv{r},\vv{r}^{\prime}+2\vv{c}},
\label{Eq:HamiltonianYBCFO b}
\\
&
J^{2,1}_{\vv{r},\vv{r}^{\prime}}\:= 
J^{\prime}_{\perp}\,
\delta^{\,}_{\vv{r},\vv{r}^{\prime}}+ J^{\,}_{\perp} \delta^{\,}_{\vv{r},\vv{r}^{\prime}-2\vv{c}},
\label{Eq:HamiltonianYBCFO c}
\\
&
J^{1,1}_{\vv{r},\vv{r}^{\prime}}=J^{2,2}_{\vv{r},\vv{r}^{\prime}}\:= 
J^{\,}_{\parallel} 
\sum_{\vv{\alpha}=\pm\vv{a^{\prime}},\pm\vv{b^{\prime}}}\, 
\delta^{\,}_{\vv{r},\vv{r}^{\prime}+\vv{\alpha}}  
.
\label{Eq:HamiltonianYBCFO d}
\end{align}
The subscript $\mathcal{L}$ for the Hamiltonian 
(\ref{Eq:HamiltonianYBCFO a}) indicates its dependence on
the choice made for the set $\mathcal{L}$ of impurity bonds.
This dependence arises from the term
\begin{equation}
H^{\,}_{\mathrm{imp}}\:=
\left(|J^{\,}_{\mathrm{imp}}|+J^{\prime}_{\perp}\right) 
\sum_{\tilde{\vv{r}}\in\mathcal{L}}  
\vv{S}^{\,}_{\tilde{\vv{r}},1} 
\cdot 
\vv{S}^{\,}_{\tilde{\vv{r}},2},
\label{Eq:HamiltonianYBCFO e}
\end{equation}
\end{subequations} 
which substitutes any ferromagnetic intra-layer bond 
associated with the exchange coupling
$J^{\prime}_{\perp}\geq0$ belonging to $\mathcal{L}$ with an
antiferromagnetic impurity bond associated with the exchange coupling
$J^{\,}_{\mathrm{imp}}\leq0$.  In the cell considered for the Monte Carlo
simulations, there are $L\times L\times L$ crystallographic unit cells,
i.e., the cell contains $N=2L^{3}$ spins. The values of the 
exchange couplings are those given in the main text 
($J^{\,}_{||}=-28.9$ meV,
$J^{\,}_{\perp}=-5.5$ meV, $J^{\prime}_{\perp}=2.7$ meV,
$J^{\,}_{\mathrm{imp}}=-95.8$ meV and $\Delta=0.5$ meV).

We employ the exchange Monte
Carlo~\cite{Hukushima:1996dn} and
overrelaxation~\cite{Alonso:1996if} method using 200 temperature
($T$) points in the range of $10 \le k^{\,}_{\mathrm{B}}T\le 30$ (meV$^{-1}$) 
for $L=28$. Replicas are exchanged between two neighbouring temperatures.  
The distribution of the temperatures is optimized so that the exchange
rate is independent of $T$.  Thermodynamic observables are measured
using the reweighting method \cite{LANDAU:2001vp}. Periodic boundary
conditions in the $ab$ plane and open boundary conditions along the
$c$ direction are used in order to deal with a magnetic spiral order
with an incommensurate wave vector.

Monte Carlo simulations are performed for several sets $\mathcal{L}$
chosen randomly. As described in the main text, the randomly chosen
impurity bonds
satisfy the constraint that there be a minimal lateral
distance $\mindist$ between impurity bonds separated by $2c$ along
the $\vv{c}$ axis of the pseudocubic lattice.  We denote with
$\langle\cdots\rangle^{\,}_{\mathrm{MC}}$ the Monte Carlo average and
with $\langle\cdots\rangle^{\,}_{\mathcal{L}}$ the average over
randomly chosen sets $\mathcal{L}$.  Denoting with 
Eq.\ (\ref{Eq:HamiltonianYBCFO aa})
the position of the unit cell, 
we define the collinear antiferromagnetic order parameter
$\mAF$ as the one that would occur in the absence of impurity bonds
\begin{equation}
m^{\,}_{\mathrm{AF}}:= 
\left\langle 
\left\langle  
\left|
\frac{1}{N}
\sum_{\substack{m,n,l\\\mu=1,2}} 
(-1)^{n+m+l}\,
\boldsymbol{S}^{\,}_{\vv{r},\mu} 
\right|
\right\rangle^{\,}_{\mathrm{MC}}
\right\rangle^{\,}_{\mathcal{L}}.
\end{equation} 
The specific heat is defined by
\begin{equation} 
C:= 
\left\langle 
\frac{
\left\langle 
\left(
H^{\,}_{\mathcal{L}}
\right)^{2} 
\right\rangle^{\,}_{\mathrm{MC}} 
- 
\left\langle 
\vphantom{\left(H^{\,}_{\mathcal{L}}\right)^{2}}
H^{\,}_{\mathcal{L}} 
\right\rangle^{2}_{\mathrm{MC}}
     }
     { 
N\,k^{\,}_{B}T^{2}} 
\right\rangle^{\,}_\mathcal{L}.
\end{equation}
The order parameter for the spiral state is defined to be 
\begin{subequations}
\begin{equation}
m^{\,}_{\mathrm{spi}}:= 
\left\langle  
\left\langle 
\left|  
\frac{1}{N-L^{2}}\,
\Pi
\right| 
\right\rangle^{\,}_{\mathrm{MC}}
\right\rangle^{\,}_\mathcal{L},
\end{equation} 
where 
\begin{equation}
\Pi:= 
\left(
\sum_{\boldsymbol{r}} 
\boldsymbol{S}^{\,}_{\boldsymbol{r},1}
\wedge
\boldsymbol{S}^{\,}_{\boldsymbol{r},2} 
+
\sum_{\boldsymbol{r}}^{1\le l \le L-1} 
\boldsymbol{S}^{\,}_{\boldsymbol{r},2} 
\wedge
\boldsymbol{S}^{\,}_{\boldsymbol{r}+\boldsymbol{z},1} 
\right) 
\cdot 
\vv{z}
\label{eq:def-P}
\end{equation}
\end{subequations}
generalizes Eq.\ (2) in the main text. 
The structure factor at 
\begin{equation}
\vv{q}=\left(\frac{1}{2},\frac{1}{2}, q^{\,}_{c}\right)
\label{eq: def q of c}
\end{equation}
is defined to be
\begin{equation} 
S(q^{\,}_{c}):=  
\frac{1}{2} 
\sum_{\mu=1,2} 
\left\langle  
\left\langle 
\left|
\frac{1}{N}   
\sum_{n,m,l} 
(-1)^{n+m}\,   
e^{-2 \pi\mathrm{i}q^{\,}_{c} l}\, 
S^{x}_{\boldsymbol{r},\mu} 
\right|^{2}
+
\left|
\frac{1}{N}   
\sum_{n,m,l} 
(-1)^{n+m}\, 
e^{-2 \pi \mathrm{i}q^{\,}_{c} l}\,  
S^{y}_{\boldsymbol{r},\mu} 
\right|^{2}
\right\rangle^{\,}_{\mathrm{MC}}
\right\rangle^{\,}_{\mathcal{L}}.
\label{eq: def spin structure factor}
\end{equation}

\section{Dependence on $\xi_{\mathrm{min}}$ and system size}

\begin{figure}
	\begin{tabular}{cc}
		\begin{minipage}{0.5\hsize}
             \includegraphics[width=\textwidth]{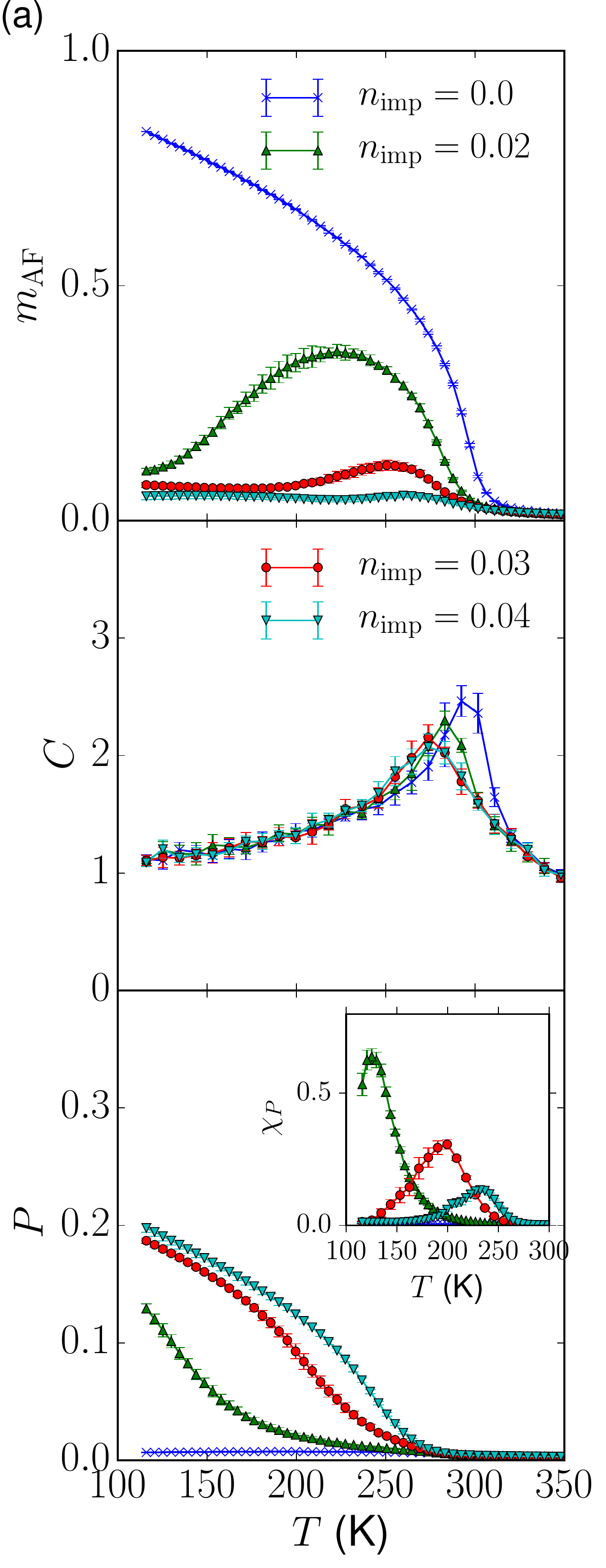}
		\end{minipage}
		&
		\begin{minipage}{0.5\hsize}
			\includegraphics[width=\textwidth]{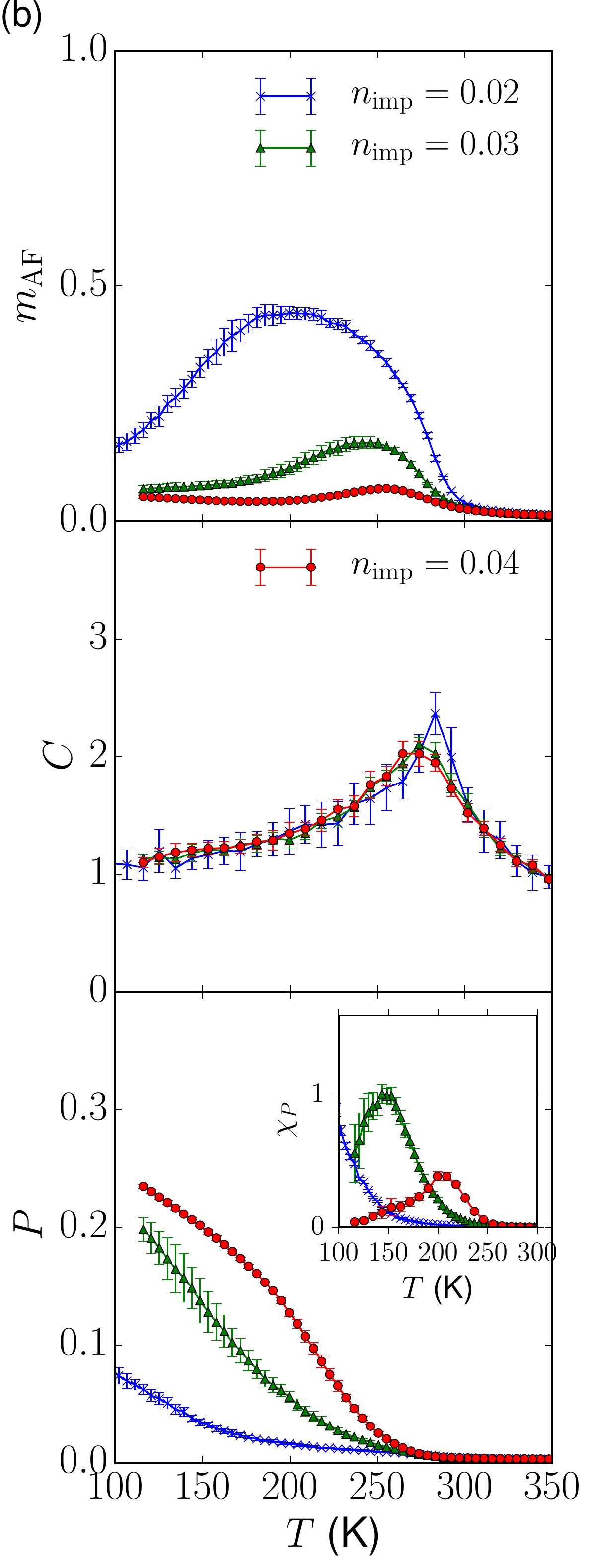}
		\end{minipage}       
	\end{tabular}
\caption{
Results of MC simulations for the temperature dependence of the collinear
order parameter $\mAF$, the specific heat $C$, the polarization $P$,
and electric susceptibility $\chi^{\,}_{\mathrm{P}}$ (inset of lowest
panels) for $\mindist=4$ (a) and $\mindist=2.5$ (b) in units of
in-plane lattice constant.
         }
\label{fig:MC-XiDep}
\end{figure}

We start by discussing the $\mindist$ dependence of the results.
Figure \ref{fig:MC-XiDep}(a) shows MC data obtained for $L=28$ at
$n^{\,}_{\mathrm{imp}}=0$, 0.02, 0.03, 0.04 for $\mindist = 4$.
Figure~\ref{fig:MC-XiDep}(b) shows the results for the case $\mindist
= 2.5$.  For both 
values of $\mindist$, the low-temperature values
of the collinear order parameter $\mAF$ (upper panel)
 are
suppressed as $n^{\,}_{\mathrm{imp}}$ is increased.  Moreover, for
$n^{\,}_{\mathrm{imp}} \neq 0$, $\mAF$ shows the re-entrant behavior
described in the main text as the temperature is decreased.
Interestingly, for the case of $\xi^{\,}_{\mathrm{min}}=2.5$ the
collinear antiferromagnetic order parameter $m^{\,}_{\mathrm{AF}}$
seems to not vanish completely at low temperatures for the lowest impurity
concentration. This might either arise from finite size effects (too
few windings of the spiral within the finite lattice) or indicate a
possible coexistence of the magnetic spiral state with a collinear
spin structure.  The middle panels of 
Figs.\,\ref{fig:MC-XiDep}(a) and \ref{fig:MC-XiDep}(b) 
show the temperature dependence of the specific heat $C$.  The peak
position in $C$ coincides well with the onset of the collinear order
parameter $\mAF$ and is approximately unaffected by
$n^{\,}_{\mathrm{imp}}$ and $\mindist$.  
The polarization $P$ (bottom panels) becomes nonvanishing below 
$T^{\,}_{\mathrm{spi}}$. For $T<T^{\,}_{\mathrm{spi}}$, $P$ increases 
continuously with decreasing $T$ to some saturation value at $T=0$.
Hence, $T^{\,}_{\mathrm{spi}}$ is interpreted as the onset of
the spiral phase through a continuous phase transition provided
$n^{\,}_{\mathrm{imp}}>0$.
The temperature below
which the magnetic spiral state appears is better identifiable by the
positions of the peaks of the dielectric susceptibility
$\chi^{\,}_{P}$, defined as
\begin{equation}
\chi^{\,}_{P}:= 
\left\langle 
\frac{
\left\langle 
\Pi^{2}
\right\rangle^{\,}_{\mathrm{MC}} 
- 
\left\langle 
|\Pi| 
\right\rangle^{2}_{\mathrm{MC}}
     }
     {
k^{\,}_{\mathrm{B}}TN
     }
\right\rangle^{\,}_{\mathcal{L}}
\label{Eq:ElectSus}
\end{equation}
and shown in the insets of the lower panels in Fig.\ \ref{fig:MC-XiDep}.

Figure \ref{fig:MC-StrFact} shows the dependence on $q^{\,}_{c}$
in Eq.\,(\ref{eq: def q of c}) of the spin-structure factor 
(\ref{eq: def spin structure factor})
computed for eight different configurations of impurities.  
The size of the system and the minimal distance are set to $L=28$
and $\mindist=4$, respectively.  
One clearly sees a peak in the dependence of 
$S(q^{\,}_{c})$ on $q^{\,}_{c}$ at a value of $q^{\,}_{c}$  that deviates
from $0.5$ as $n^{\,}_{\mathrm{imp}}$ is increased. This deviation 
approximately increases linearly with the concentration 
of impurity bonds $n^{\,}_{\mathrm{imp}}$,
in agreement with the predictions of the low-temperature effective model.
\begin{figure}
\centering\includegraphics[width=0.65\textwidth]{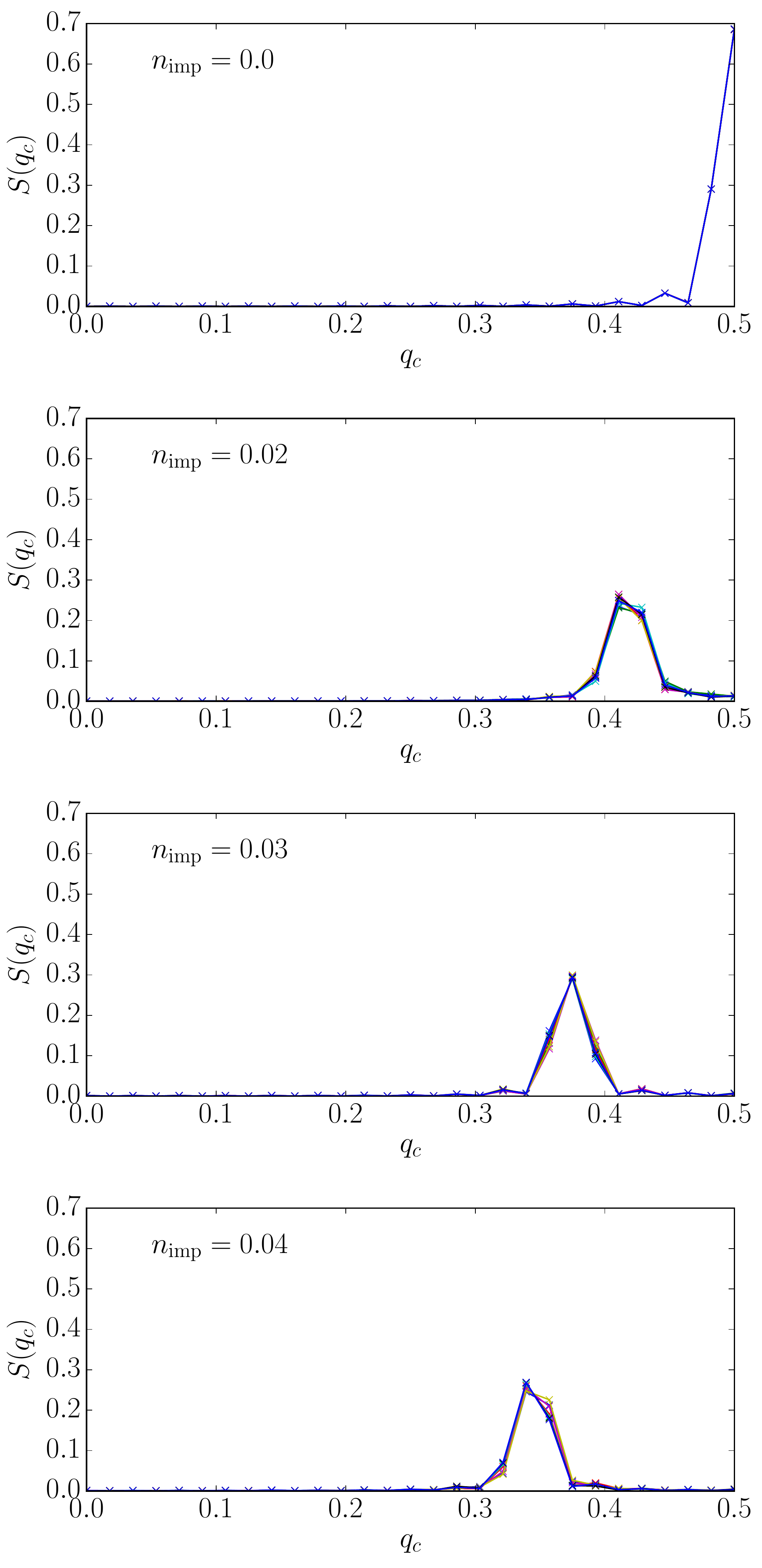}
\caption{
Spin structure factor $S(q^{\,}_{c})$ computed for different eight
configurations of impurity bonds with $L=28$ and $\mindist=4$.  
        }
\label{fig:MC-StrFact}
\end{figure}

\begin{figure}
\includegraphics[width=0.45\textwidth]{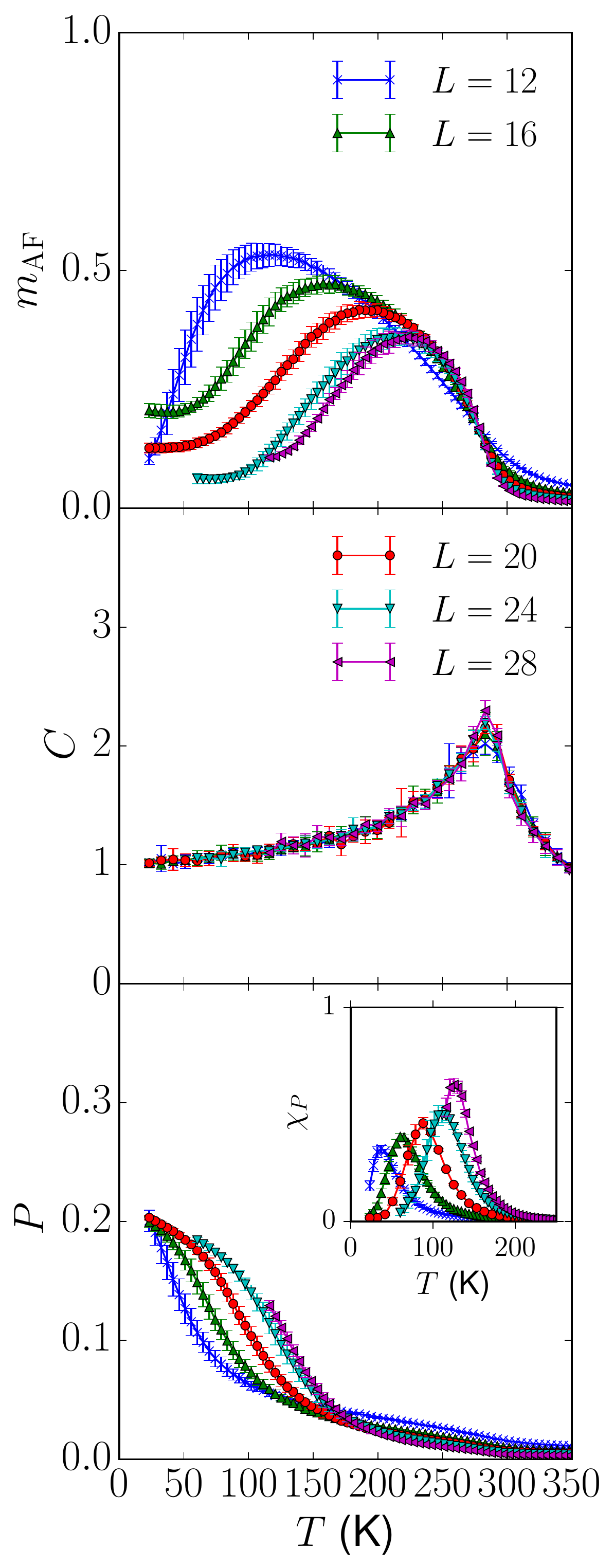}
\caption{
MC results for the temperature dependence of the collinear order
parameter $\mAF$, the specific heat $C$, the polarization $P$, and
electric susceptibility $\chi^{\,}_{P}$ (inset) 
for various sizes of the lattice. 
The minimal distance is fixed at $\mindist=4$,
while the concentration of impurity bonds is
$\ximp=0.02$.  
        }
\label{fig:MC-SizeDep}
\end{figure}

To study the approach to the thermodynamic limit, 
we investigate the system-size dependence of the
MC results for $n^{\,}_{\mathrm{imp}}=0.02$.  As shown in
Fig.\ \ref{fig:MC-SizeDep}, the dependence of $\mAF$ on the system
size is not monotonic around 120 K: $\mAF$ has a small but
non-vanishing residual value even at $L=28$.  Therefore, as the system
size is not large enough to obtain convergent results for $\mAF$, we
cannot exclude the possibility of the coexistence of the spin
collinear order and the spin spiral order at low temperatures.  On the
other hand, $P$ converges with increasing $L$
rapidly for the lowest temperature
considered to a nonvanishing and finite value.
For example, at $T\simeq 23$ K, $P$
does not change appreciably when increasing $L$ from $L=12$ to $L=28$.  
This observation establishes the
existence of the spin-spiral state at low $T$.  However, the peak
position in $\chi^{\,}_{P}$ drifts toward high temperatures as the
system size is increased. This indicates that our calculations might
actually underestimate the value of $T^{\,}_{\mathrm{spi}}$.  Figure
\ref{fig:MC-TFiniteSize} shows the system-size dependence of $\TN$
(peak position in $C$) and $\Ts$ (peak position in $\chi^{\,}_{P}$).
This clearly shows that $\Ts$ is still monotonically increasing 
for $L$ increasing from 12 to 28, 
while $T^{\,}_{N}$ is approximately constant 
for $L$ increasing from 12 to 28.

\begin{figure}
\centering\includegraphics[width=0.5\textwidth]{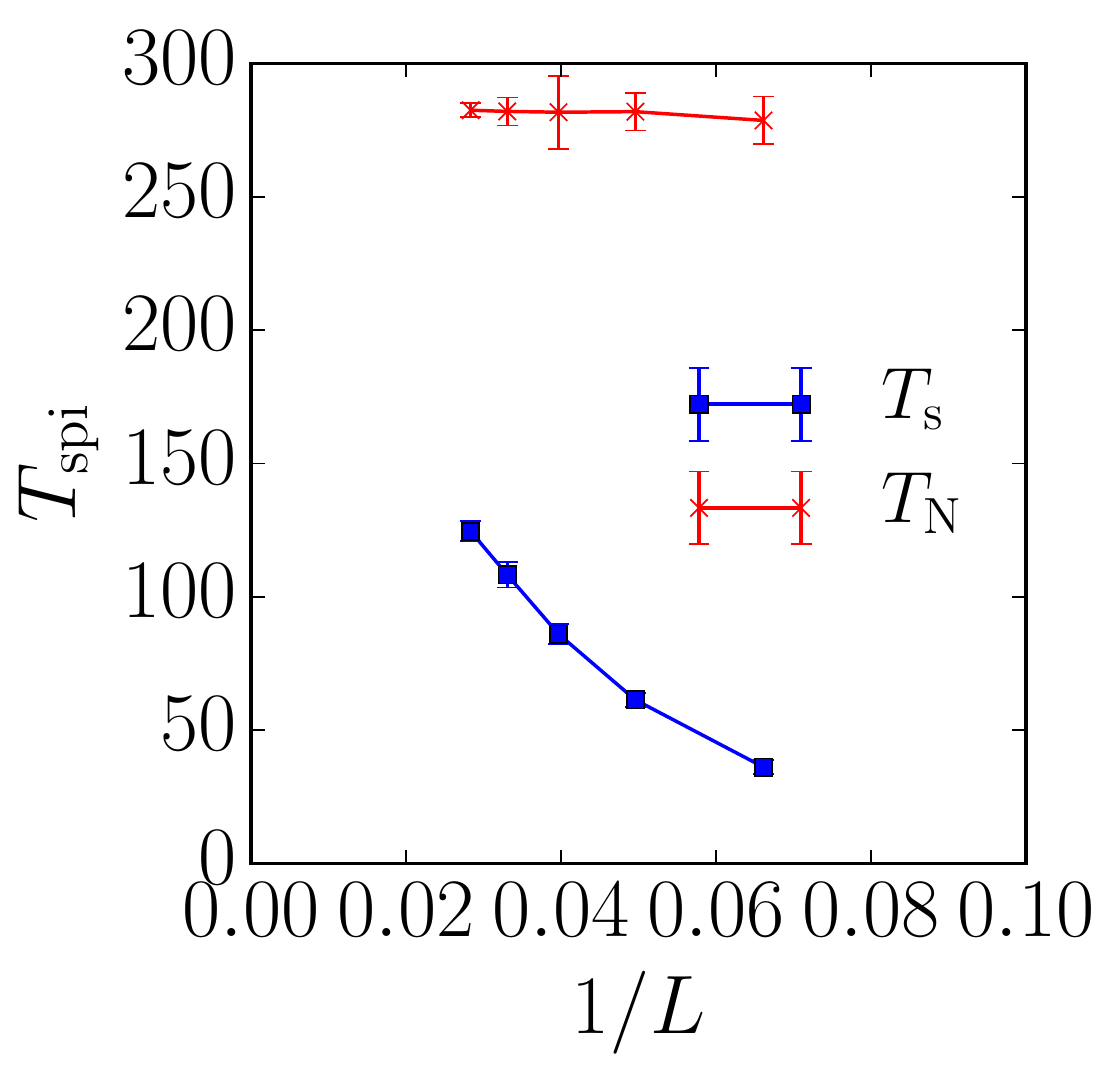}
\caption{
System-size dependence of the peak positions in the specific heat $C$
($\TN$) and the susceptibility $\chi^{\,}_{P}$ ($\Ts$) at $\ximp=0.02$
and $\mindist=4$.  
        }
\label{fig:MC-TFiniteSize}
\end{figure}

Although our MC results for given large $L\times L\times L$ strongly
suggest the existence of a finite $\Ts$, extrapolating $\Ts$ in the
thermodynamic limit $L\to\infty$ with $n^{\,}_{\mathrm{imp}}$ held
fixed is difficult.  Furthermore, the transition may be sensitive to
the geometry of the system since the effective interactions between
impurity bonds are long ranged.  Thus, we systematically study the
geometry dependence of the spin-spiral transition.
Figure \ref{fig:MC-geom} compares $\Ts$ obtained for different
geometries for $n^{\,}_{\mathrm{imp}}$=0.02 and $\mindist$=4.  
We adopt tetragonal shapes by which the dimensions
$L\times L\times L$
are substituded by the dimensions
$L^{\,}_{a}\times L^{\,}_{b}\times L^{\,}_{c}$
and we vary the ratio
$L^{\,}_{a}/L^{\,}_{c}=L^{\,}_{b}/L^{\,}_{c}$ from 1 to 8. As the ratio
$L^{\,}_{a}/L^{\,}_{c}$ becomes larger, $\Ts$ tends to increase for the
same numbers of spins $N^{\,}$.  This trend is consistent with the
dipole effective interaction, which is ferromagnetic in transverse
directions $\vv{a}^{\prime}$ and $\vv{b}^{\prime}$.  All the series
of $\Ts$ for different values of the $L^{\,}_{a}/L^{\,}_{c}$ ratio 
extrapolate to $\Ts\simeq 200$ - $250$ K in the thermodynamic
limit.

To confirm this result, 
we present the MC simulation for the Binder cumulant $g^{\,}_{P}$ of the
the polarization $P$,
\begin{equation}
g^{\,}_{P}:= 
\left\langle
\frac{1}{2}
\left(
3
-
\frac{
\langle \Pi^{4} \rangle^{\,}_{\mathrm{MC}}
     }
     {
\langle \Pi^{2}\rangle^{2}_{\mathrm{MC}}
     }
\right)
\right\rangle^{\,}_{\mathcal{L}}, 
\end{equation}
where $\Pi$ is defined by Eq.\ (\ref{eq:def-P}).
The Binder cumulant is first computed for each configuration and then
averaged over different configurations of impurity bonds.  We show the
temperature dependence of $g^{\,}_{P}$ for various values of the ratio
$L^{\,}_{a}/L^{\,}_{c}$ in Fig.\ \ref{fig:MC-binder}.  
One clearly sees an intersection of the data at $\Ts\simeq 225$ K, 
which is lower than $\Tc$.  
The intersection of the Binder cumulants locates the value of
$T^{\,}_{\mathrm{spi}}$ at $\Ts\simeq 225$ K $<$ $T^{\,}_{\mathrm{N}}
\simeq 275$ K.

\begin{figure}
\centering\includegraphics[width=0.75\textwidth]{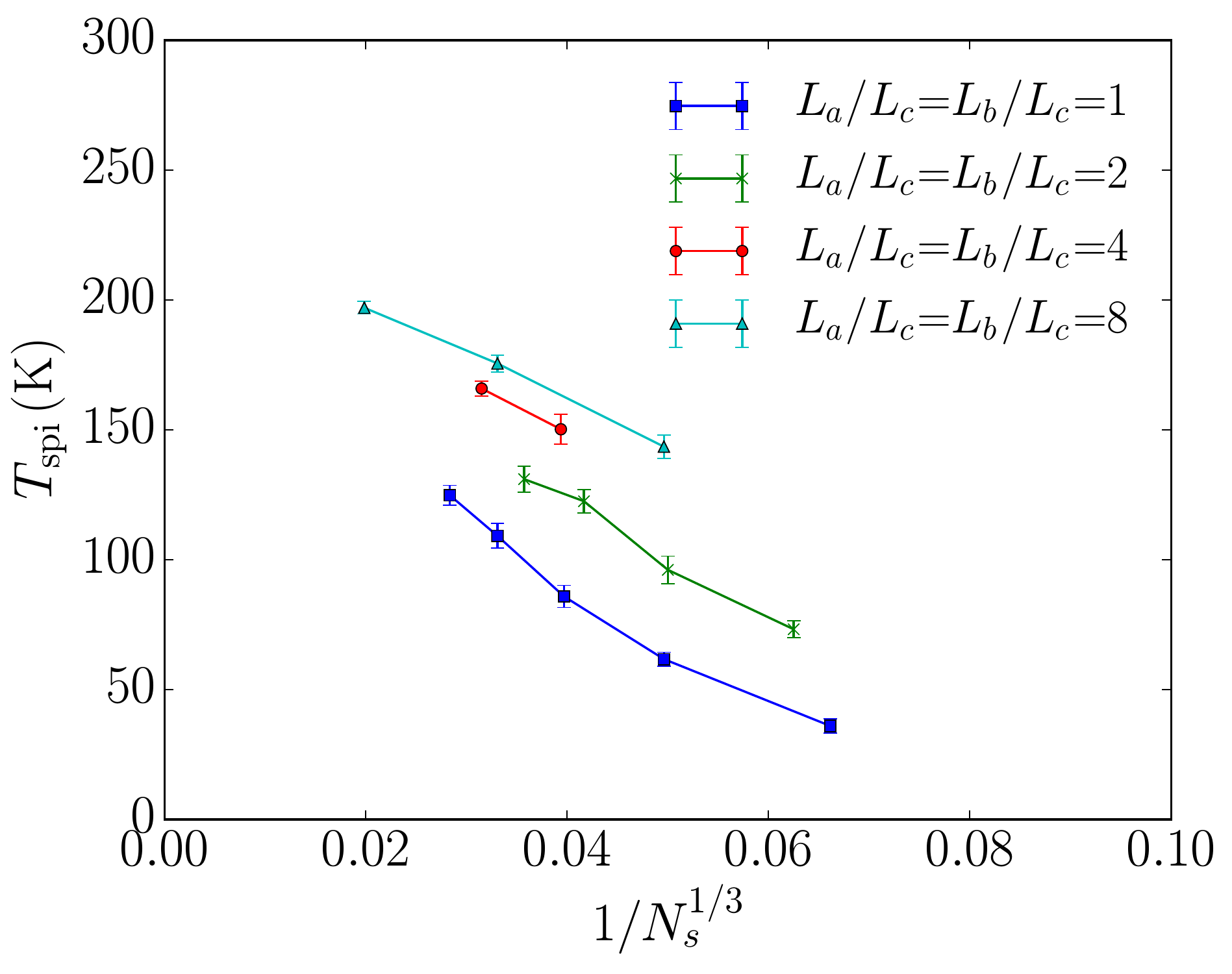}
\caption{
System-size dependence of the peak position in $\chi^{\,}_{P}$
for different geometries. The MC data were obtained for
$\ximp$=0.02 and $\mindist$=4.  
        }
\label{fig:MC-geom}
\end{figure}

\begin{figure}
\centering\includegraphics[width=0.75\textwidth]{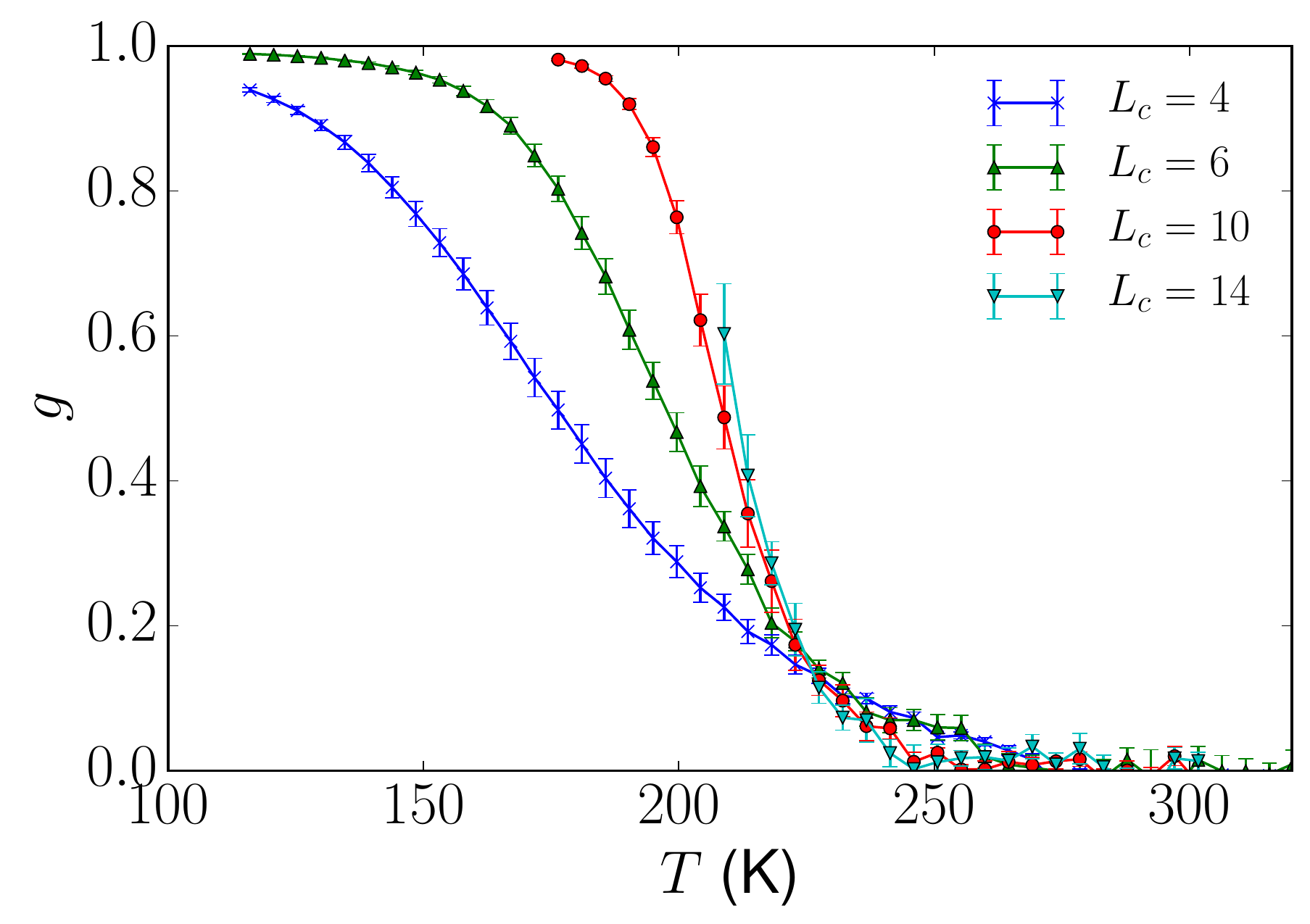}
\caption{
Binder cumulant of the polarization $P$ computed for $\ximp$=0.02 and
$\mindist$=4 with system sizes of 
$(8 L^{\,}_{c}, 8 L^{\,}_{c}, L^{\,}_{c})$.  
        }
\label{fig:MC-binder}
\end{figure}


\end{document}